\documentclass[aps,prb,superscriptaddress]{revtex4}
\usepackage{amsmath,amssymb,mathrsfs}
\usepackage[dvips]{graphicx}
\usepackage{psfrag}
\usepackage{color}

\newcommand{\bs}[1]{\boldsymbol{#1}}
\newcommand{\up}{\uparrow}
\newcommand{\dw}{\downarrow}
\newcommand{\comm}[2]{\left[#1,#2\right]}
\newcommand{\anticomm}[2]{\left\{#1,#2\right\}}
\newcommand{\ket}[1]{\left|#1\right\rangle}
\newcommand{\bra}[1]{\left\langle#1\right|}
\newcommand{\sgn}[1]{\text{sgn}(#1)\,}
\newcommand{\ii}{\text{i}}

\def\prl#1#2#3{Phys.\ Rev.\ Lett.\ {\bf #1}, #2 (#3)}

\def\prb#1#2#3{Phys.\ Rev.\ B {\bf #1}, #2 (#3)}

\def\BSCCO{Bi$_2$Sr$_2$CaCu$_2$O$_{8+\delta}$}
\def\C60{A$_x$C$_{60}$}

\def\HgCu3{HgCa$_2$Cu$_3$O$_{8+y}$}
\def\HgCu4{HgBa$_2$Ca$_3$Cu$_4$O$_{10+y}$}
\def\TlCu{Tl$_2$Ba$_2$CuO$_{6+\delta}$}
\def\TlCu3{Tl$_2$Ba$_2$Ca$_2$Cu$_3$O$_{10+y}$}
\def\TlCu4{Tl$_2$Ba$_2$Ca$_3$Cu$_4$O$_{12+y}$}

\def\BiCu3{Bi$_2$Sr$_2$Ca$_{2}$Cu$_3$O$_y$}

\def\8LSCO{La$_{1.88}$Sr$_{.12}$CuO$_4$}
\def\110LNSCO{La$_{1.5}$Nd$_{0.4}$Sr$_{0.1}$CuO$_{4}$}
\def\stage4LCO{La$_{2}$CuO$_{4+\delta}$}
\def\Y248{YBa$_2$Cu$_4$O$_8$}

\def\hts{high temperature superconductors}

\def\NbSe2{NbSe$_2$}
\def\TaSe2{TaSe$_2$}
\def\TiSe2{TiSe$_2$}
\def\NaCoOH2O{Na$_{0.3}$CoO$_{2y}$H$_2$O}
\def\MgB2{MgB${}_2$}

\newcommand{\be}{\begin{equation}}
\newcommand{\ee}{\end{equation}}
\newcommand{\bea}{\begin{eqnarray}}
\newcommand{\eea}{\end{eqnarray}}

\def\nn{\nonumber\\}
\def\fr#1{(\ref{#1})}

\allowdisplaybreaks[4]

\begin{document}
\title{Boundary effects on the local density of states of one-dimensional\\ 
  Mott insulators and charge density wave states}
\author{Dirk Schuricht}
\affiliation{Institute for Theory of Statistical Physics, 
RWTH Aachen, 52056 Aachen, Germany}
\affiliation{JARA-Fundamentals of Future Information Technology}
\author{Fabian H. L. Essler}
\affiliation{The Rudolf Peierls Centre for Theoretical Physics, 
University of Oxford, 1 Keble Road, OX1 3NP, Oxford, UK}
\author{Akbar Jaefari}
\affiliation{Department of Physics, University of Illinois at Urbana-Champaign,
1110 W. Green St, Urbana, Illinois 61801-3080, USA}
\author{Eduardo Fradkin}
\affiliation{Department of Physics, University of Illinois at Urbana-Champaign,
1110 W. Green St, Urbana, Illinois 61801-3080, USA}
\date{\today}
\pagestyle{plain}

\begin{abstract}
  We determine the local density of states (LDOS) for spin-gapped
  one-dimensional charge density wave (CDW) states and Mott insulators in the
  presence of a hard-wall boundary. We calculate the boundary contribution to
  the single-particle Green function in the low-energy limit using field
  theory techniques and analyze it in terms of its Fourier transform in both
  time and space. The boundary LDOS in the CDW case exhibits a singularity at
  momentum $2k_\mathrm{F}$, which is indicative of the pinning of the CDW
  order at the impurity. We further observe several dispersing features at
  frequencies above the spin gap, which provide a characteristic signature of
  spin-charge separation. This demonstrates that the boundary LDOS can be used
  to infer properties of the underlying {\it bulk} system. In presence of a
  boundary magnetic field mid-gap states localized at the boundary emerge. We
  investigate the signature of such bound states in the LDOS.  We discuss
  implications of our results on STM experiments on quasi-1D systems such as
  two-leg ladder materials like Sr$_{14}$Cu$_{24}$O$_{41}$. By exchanging the
  roles of charge and spin sectors, all our results directly carry over to the
  case of one-dimensional Mott insulators.
\end{abstract}
\maketitle

\section{Introduction}
Scanning tunneling microscopy (STM) and spectroscopy (STS) methods
have proved to be a useful tool for studying strongly correlated
electron systems such as carbon nanotubes~\cite{Odom-02},
{\hts}
(HTSC)~\cite{Howald-01,Hoffman-02,Kohsaka-07,Fischer-07}
and rare-earth compounds~\cite{Fang-07}. 
STM experiments measure the tunneling current $I$ between the sample
and the STM tip as a function of its position $x$ and the applied
voltage $V$.  This current can be expressed in terms of the local
densities of states (LDOS) in the sample $N(E,x)$ and the tip
$N_\mathrm{tip}(E)$ as~\cite{Fischer-07} 
\begin{equation}
I(V,x)\propto\int dE\,\bigl[f(E-eV)-f(E)\bigr]\,
N_\mathrm{tip}(E-eV)\,N(E,x),
\label{eq:STM1}
\end{equation}
where $f(E)$ denotes the Fermi function. Assuming a structureless density
of states in the tip, $N_\mathrm{tip}=\text{const}$, this gives the
following expression for the local tunneling conductance
\begin{equation}
\frac{dI(V,x)}{dV}\propto\int dE\,f'(E-eV)\,N(E,x).
\label{eq:defconductance}
\end{equation}
Eq.~\fr{eq:defconductance} shows that the tunneling conductance is
proportional to the thermally smeared $N(E,x)$ of the sample at the
position of the tip. A non-trivial spatial dependence of the LDOS
arises in presence of impurities. These break translational invariance
and lead to a modification of the LDOS in their vicinity, from which
one can infer characteristic properties of the \emph{bulk} state of
matter as well as the nature of its electronic excitations. Spatial
modulations of the LDOS can be analyzed in terms the Fourier transform
of the tunneling conductance. It follows from
\eqref{eq:defconductance} that this quantity is directly proportional
to the corresponding Fourier transform of the LDOS, $N(E,Q)$. This
method of analyzing STM data was used very
successfully~\cite{Hoffman-02} to study quasiparticle interference in
Bi$_2$Sr$_2$CaCu$_2$O$_{8+\delta}$. The spin dependence of the LDOS
has also been investigated using magnetic tips\cite{Wiesendanger-90}.
Theoretical studies of STS have focused in particular on Luttinger
liquids\cite{Eggert-00,Kivelson-03} and
HTSC\cite{Kivelson-03,Polkovnikov-02,Podolsky-03}.  In the Luttinger liquid case
an impurity has the same effects at low energies as a physical
boundary\cite{KaneFisher92prl}, which motivated studies of the LDOS in
the vicinity of a chain end. The case of strongly correlated
one-dimensional (1D) systems with spin or charge gaps is of
considerable interest as well and pertains to quasi 1D charge density
wave (CDW) systems\cite{1DCDW} and Mott
insulators\cite{1DMott,Bechgaard}, carbon nanotubes\cite{Wildoer-98},
(doped) two-leg ladder materials\cite{Tennant,Thierry}, and the stripe phases
of HTSC\cite{Arrigoni-04}.  As compared to the Luttinger liquid case
the presence of an interaction-induced gap makes these problems much
more difficult to treat theoretically. In the following we will
determine the LDOS for the low-energy limit of 1D CDW states and Mott
insulators in presence of a single boundary. The latter can be thought
of as arising as the result of the presence of a strong potential
impurity. Alternatively one can imagine inducing a boundary in a two
tip STM setup, where the first tip is used to induce a boundary by
applying a high voltage and the LDOS is then measured with a second
tip. A short summary of our results has appeared previously\cite{prl}.

The outline of this paper is as follows: in section \ref{sec:model}
we present the field theory limit of 1D CDW states and Mott insulators
in presence of a single boundary. In section \ref{sec:GF} we summarize
our results for the single-particle Green function. These are then
used to determine the Fourier transform $N_\sigma(E,Q)$ of the local
density of states for hard-wall boundary conditions in section
\ref{sec:LDOS}. Signatures of spin and charge excitations visible in
$N_\sigma(E,Q)$ are discussed in some detail. The effects of more
general boundary conditions, including the formation of boundary
bound states, are described in section
\ref{sec:generalBC}. Section \ref{sec:finitetemp} deals with the
effects of finite temperatures and implications of our results for STM
experiments are discussed in section \ref{sec:exp}. The technical
details of our calculations are presented in several appendices.

\section{The model}
\label{sec:model}
Our analysis of the LDOS is based on the continuum description of certain 1D
CDW states and Mott insulators. The resulting quantum field theory for both
cases is known as the U(1) Thirring model\cite{U1Thirring} (with two
``flavors''.)  The latter is known to arise as the effective low-energy
description of a number of lattice models of spin-$1/2$ electrons as we
discuss next.
\begin{enumerate}
\item{} Half-filled repulsive Hubbard model \cite{book}:

This is the standard model for single band 1D Mott insulators. The
Hamiltonian is of the form
\bea
H&=&-t\sum_{j,\sigma}
\left[c^\dagger_{j,\sigma}c_{j+1,\sigma}+
c^\dagger_{j+1,\sigma}c_{j,\sigma}\right]
+U\sum_j \left(n_{j,\uparrow}-\frac{1}{2}\right)
\left(n_{j,\downarrow}-\frac{1}{2}\right),
\label{HHub}
\eea
where $n_{j,\sigma}=c^\dagger_{j,\sigma}c_{j,\sigma}$ and
$n_j=n_{j,\uparrow}+n_{j,\downarrow}$ are electron number operators
and $U>0$.

\item{} 1D Holstein model \cite{Holstein59a}:

The Holstein model provides an example of an incommensurate CDW state and
describes a partially filled band of spin-$1/2$ electrons coupled to
dispersionless phonons of frequency $\omega_0=\sqrt{\frac{k}{M}}$ 
\bea
H=-t\sum_{j,\sigma}\left[c^\dagger_{j,\sigma}c_{j+1,\sigma}+
c^\dagger_{j+1,\sigma}c_{j,\sigma}\right]
+\sum_j\left[\frac{P_j^2}{2M}+\frac{k}{2}Q_j^2\right]
-\lambda\sum_{j,\sigma}Q_jn_{j,\sigma}.
\label{HHolstein}
\eea
Integrating out the phonons induces a retarded attractive
electron-electron interaction. In the limit $t\ll\omega_0$ 
the retardation effects can be neglected, leading to an effective
{\sl attractive} Hubbard model with $U\propto -\lambda^2$. 

\item{} Su-Schrieffer-Heeger model \cite{Heeger-88}:

A second example of an incommensurate CDW state is provided by the
Su-Schrieffer-Heeger model, which describes a partially filled band of
spin-$1/2$ electrons coupled to dispersing phonons
\bea
H=-\sum_{j,\sigma}\bigl[t-\lambda(Q_{j+1}-Q_j)\bigr]\left[
c^\dagger_{j,\sigma}c_{j+1,\sigma}+c^\dagger_{j+1,\sigma}c_{j,\sigma}\right]
+\sum_j\left[\frac{P_j^2}{2M}+\frac{k}{2}\left[Q_{j+1}-Q_j\right]^2\right].
\label{HSSH}
\eea Taking the continuum limit of \fr{HSSH} (which describes the behavior at
low frequencies $\omega\alt t$) it was shown by Fradkin and Hirsch
\cite{FradkinHirsch83} that the regime of large phonon frequencies
$t\ll\omega_0=\sqrt{\frac{k}{M}}$ is described by the U(1) Thirring
model\cite{U1Thirring,caveat}.
\end{enumerate}
In all three cases a continuum description of the low-energy
electronic degrees of freedom is obtained by considering only the modes
in the vicinity of the Fermi points $\pm k_\mathrm{F}$. 
 The
lattice electron annihilation operators are expressed in
terms of slowly varying right- and left-moving Fermi fields as 
\begin{equation}
\label{eq:lowenergy}
\frac{c_{j,\sigma}}{\sqrt{a_0}}\rightarrow\Psi_{\sigma}(x)=
e^{\ii k_\mathrm{F} x} R_\sigma(x)+
e^{-\ii k_\mathrm{F} x} L_\sigma(x),
\end{equation}
where $a_0$ is the lattice spacing, $x=ja_0$, and $\sigma=\up,\dw$ labels the
spin. In the bulk the fields $R_\sigma$ and $L_\sigma$ are bosonized
according to
\begin{eqnarray}
R^\dagger_\sigma(\tau,x)&=&\frac{\eta_\sigma}{\sqrt{2\pi}}\,
\exp\Bigl(\tfrac{\ii}{2}\phi_\mathrm{c}(\tau,x)\Bigr)\,
\exp\Bigl(\tfrac{\ii}{2}f_\sigma\phi_\mathrm{s}(\tau,x)\Bigr),
\label{eq:bosonizationR}\\*
L^\dagger_\sigma(\tau,x)&=&\frac{\eta_\sigma}{\sqrt{2\pi}}\,
\exp\Bigl(-\tfrac{\ii}{2}\bar{\phi}_\mathrm{c}(\tau,x)\Bigr)\,
\exp\Bigl(-\tfrac{\ii}{2}f_\sigma\bar{\phi}_\mathrm{s}(\tau,x)\Bigr),
\label{eq:bosonizationL}
\end{eqnarray}
where the Klein factors $\eta_\sigma$ satisfy anticommutation rules
$\anticomm{\eta_\sigma}{\eta_\sigma'}=2\delta_{\sigma\sigma'}$ and
$f_\up=1=-f_\dw$. The fields $\phi_a$ and $\bar{\phi}_a$
are the chiral components of the canonical Bose fields $\Phi_a$ and
their dual fields $\Theta_a$,
\begin{equation}
\label{eq:gaussfields}
\Phi_a=\phi_a+\bar{\phi}_a,\; \Theta_a=\phi_a-\bar{\phi}_a,\quad
a=\mathrm{c},\mathrm{s}.
\end{equation}
In the bulk the Hamiltonian density then can be cast in the
spin-charge separated form
\bea
{\cal H}(x)&=&\sum_{a=\mathrm{c,s}}{\cal H}_{a}(x),\nn*
{\cal H}_{\rm a}&=&\frac{v_a}{16\pi}
\biggl[\frac{1}{K_a^2}\bigr(\partial_x\Phi_a\bigr)^2+
K_a^2\bigr(\partial_x\Theta_a\bigr)^2\biggr]
-\frac{g_a}{(2\pi)^2}\ \cos\Phi_a,
\eea
The charge and spin velocities $v_\mathrm{c,s}$, the Luttinger parameters
$K_\mathrm{c,s}$ and coupling constants $g_\mathrm{c,s}$ are functions of the
hopping integrals and interaction strengths defining the underlying
microscopic model. The cases discussed above correspond to the following
parameter regimes: 
\begin{enumerate}
\item{} Mott insulators

As a result of repulsive electron-electron interactions we have
$v_\mathrm{c}>v_\mathrm{s}$ and $K_\mathrm{c}<1$. The $\cos\Phi_\mathrm{c}$
perturbation in the charge sector is relevant and opens up a gap. 
The $\cos\Phi_\mathrm{s}$ interaction in the spin sector is 
marginally irrelevant and flows to zero under the renormalization
group. We therefore neglect it in the following.
\item{} Electron-Phonon Systems

At low energies the electron-phonon coupling induces an attractive
electron-electron interaction, which results in
$v_\mathrm{s}>v_\mathrm{c}$ and $K_\mathrm{c}>1>K_{\rm s}$. The
$\cos\Phi_\mathrm{c}$ term is irrelevant, while $\cos\Phi_\mathrm{s}$
term is relevant (marginally relevant in the spin-SU(2) symmetric
case $K_\mathrm{s}=1$) and opens up a gap in the spin sector.
\end{enumerate}
In both cases we end up with a spin-charge separated theory of a
gapless Luttinger liquid and a sine-Gordon model.

We now imagine a strong, local potential to be present. It is well known from
the work of Kane and Fisher\cite{KaneFisher92prl} that the coupling to the
impurity is {\em relevant}, leading to a crossover at a characteristic
dynamical energy (and temperature) scale called ``$T_\mathrm{K}$'', below
which the system is effectively cut into two disconnected parts, and to a
pinning of the CDW.  This potential could be due to a strong potential
impurity or an STM tip. We model the strong impurity potential by a boundary
condition on the continuum electron field 
\begin{equation}
\Psi_\sigma(x=0)=0.  
\end{equation}
An important physical consequence of the pinning of the CDW at the impurity
(or boundary) is the development of an {\em induced static CDW order} in this
(effectively) quantum critical system. This induced static CDW order, often
referred to\cite{white-2002} as a ``Friedel oscillation'', leads to
non-dispersive features in the LDOS\cite{Kivelson-03} which can be detected in
STM and STS experiments.

Our effective low-energy Hamiltonian in the CDW case 
(in the case of a Mott insulator the roles of spin and charge sectors are
interchanged) then becomes
\begin{eqnarray}
H&=&\sum_{a=\rm c,s}H_a,\label{eq:hamiltonian}\\
H_\mathrm{c}&=&\frac{v_\mathrm{c}}{16\pi}\int_{-\infty}^0 dx 
\biggl[\frac{1}{K_\mathrm{c}^2}\bigr(\partial_x\Phi_\mathrm{c}\bigr)^2+
K_\mathrm{c}^2\bigr(\partial_x\Theta_\mathrm{c}\bigr)^2\biggr],
\label{eq:chargehamiltonian}\\
H_\mathrm{s}&=&\frac{v_\mathrm{s}}{16\pi}\int_{-\infty}^0 dx 
\biggl[\frac{1}{K_\mathrm{s}^2}\bigr(\partial_x\Phi_\mathrm{s}\bigr)^2+
K_\mathrm{s}^2\bigr(\partial_x\Theta_\mathrm{s}\bigr)^2\biggr]
-\frac{g_\mathrm{s}}{(2\pi)^2}\int_{-\infty}^0 dx\,
\cos\Phi_\mathrm{s},
\label{eq:spinhamiltonian}
\end{eqnarray}
where the Bose fields are subject to the hard-wall boundary conditions
(we consider more general boundary conditions in Sec.~\ref{sec:generalBC})
\be
\Phi_\mathrm{c,s}(x=0)=0.
\label{hardwall}
\ee In Appendix~\ref{sec:RG} we argue that a weak potential impurity
renormalizes to strong coupling even for moderate attractive
interactions, suggesting that this situation too can be modeled in
terms of the boundary conditions \fr{hardwall}.  We note that our
starting point \fr{eq:hamiltonian}--\fr{hardwall} differs from the
model considered in Ref.~\onlinecite{Tsvelik08}, where the impurity
couples only to the gapless charge sector.

The charge sector (\ref{eq:chargehamiltonian}) describes gapless
collective charge excitations propagating with velocity
$v_\mathrm{c}$, which carry charge $\mp e$ and are commonly referred to
as holons and antiholons respectively. On the other hand, the spin
excitations or spinons are described by the sine-Gordon model on the
half-line (\ref{eq:spinhamiltonian}), which is known to be integrable
for quite general boundary conditions~\cite{GhoshalZamolodchikov94,MacIntyre95}.
In the regime $K_\mathrm{s}>1/\sqrt{2}$ the elementary bulk
excitations are gapped solitons and antisolitons which correspond to
up- and down-spin spinons respectively. For $K_\mathrm{s}<1/\sqrt{2}$
propagating breather (soliton-antisoliton) bound states occur as well.
At the Luther-Emery point (LEP) $K_\mathrm{s}=1/\sqrt{2}$ the spin
sector is equivalent to a free massive Dirac
fermion~\cite{LutherEmery74}. The exact bulk scattering matrix was
first derived by Zamolodchikov~\cite{Zamolodchikov77}; the boundary
reflection matrices of solitons and
antisolitons~\cite{GhoshalZamolodchikov94} and
breathers~\cite{Ghoshal94} were derived by Ghoshal and Zamolodchikov.
We will restrict ourselves to the regime $K_\mathrm{s}\ge 1/\sqrt{2}$
throughout, which implies that no breathers exist.

The lattice models discussed above give rise to the simplest kind
of one dimensional CDW state/Mott insulator. More complicated versions
arise in strongly correlated two- and three-leg ladder
systems~\cite{White-94,Arrigoni-04}. 
In these systems, even though electron interactions are strongly
repulsive, a Mott state with a finite (and typically large) spin gap
is found for a range of dopings close to half filling. While the
precise description of the spin sector for two-leg ladders is
considerably more complicated \cite{soN}, we expect our calculation to
capture important qualitative features.

\section{Green function}
\label{sec:GF}
The central object of our study is the time-ordered Green
function in Euclidean space,
\begin{equation}
\label{eq:GF}
G_{\sigma\sigma'}(\tau,x_1,x_2)=
-\bra{0_\mathrm{b}}\mathcal{T}_\tau\,\Psi_\sigma(\tau,x_1)\,
\Psi_{\sigma'}^\dagger(0,x_2)\ket{0_\mathrm{b}},
\end{equation}
where $\ket{0_\mathrm{b}}$ is the ground state of \eqref{eq:hamiltonian} in
the presence of the boundary and $\tau=\ii t$ denotes imaginary time. The spin
takes the values $\up$ and $\dw$. At low energies the linearization around the
Fermi points yields the decomposition
\begin{equation}
\label{eq:GFlowenergy}
G_{\sigma\sigma'}=
e^{\ii k_\mathrm{F}(x_1-x_2)}\,G^{RR}_{\sigma\sigma'}
+e^{-\ii k_\mathrm{F}(x_1-x_2)}\,G^{LL}_{\sigma\sigma'}
+e^{\ii k_\mathrm{F}(x_1+x_2)}\,G^{RL}_{\sigma\sigma'}
+e^{-\ii k_\mathrm{F}(x_1+x_2)}\,G^{LR}_{\sigma\sigma'},
\end{equation}
where e.g. $G^{RL}_{\sigma\sigma'}=-\bra{0_\mathrm{b}}\mathcal{T}_\tau\,
R_\sigma(\tau,x_1)\,L^\dagger_{\sigma'}(0,x_2)\ket{0_\mathrm{b}}$.  As we are
interested in the LDOS, we ultimately want to set $x_1=x_2$.  Below we will
calculate the spatial Fourier transform of the LDOS as physical properties can
be more easily identified. In momentum space the $RL$ and $LR$ contributions
occur in a different region ($Q\approx \pm 2k_\mathrm{F}$) compared to the
$RR$ and $LL$ parts ($Q\approx 0$). In absence of a boundary we have
$G^{RL}_{\sigma\sigma'}=G^{LR}_{\sigma\sigma'}=0$ as the charge parts of these
Green functions vanish. In presence of a boundary left and right sectors are
coupled and the Fourier transform of the Green function
(\ref{eq:GFlowenergy}) concomitantly acquires a nonzero component at $Q\approx
\pm 2k_\mathrm{F}$, which provides a particularly clean way of investigating
boundary effects. For this reason we first focus on the $2k_\mathrm{F}$-part
of the Green function and study the small momentum regime afterwards.

The Green function $G_{\sigma\sigma'}^{RL}$ factorizes into a
product of correlation functions in the spin and charge sectors. The
charge part can be determined by standard
methods~\cite{FabrizioGogolin95,Cardy84,DiFrancescoMathieuSenechal97}
(see App.~\ref{sec:appgfcs}).  On the other hand, the integrability of
the sine-Gordon model on the half-line \eqref{eq:spinhamiltonian}
enables us to calculate correlation functions in the spin sector using
the boundary state formalism introduced by Ghoshal and
Zamolodchikov~\cite{GhoshalZamolodchikov94} together with a
form-factor
expansion~\cite{Smirnov92book,Fring-93,Lukyanov95,LukyanovZamolodchikov01,Delfino04,EsslerKonik05}.
As we show in App.~\ref{sec:appgfss} the leading terms of this
expansion yield ($\tau>0$, $x_1<x_2$)
\begin{eqnarray}
G^{RL}_{\sigma\sigma'}(\tau,x_1,x_2)&=&
g_\mathrm{c}(\tau,x_1,x_2)\,g_\mathrm{s}(\tau,x_1,x_2),\label{eq:GRL}\\
g_\mathrm{c}(\tau,x_1,x_2)&=&-\frac{\delta_{\sigma\sigma'}}{2\pi}
\frac{1}{\bigl(v_\mathrm{c}\tau-2\ii R\bigr)^a}\,
\frac{1}{\bigl(v_\mathrm{c}\tau+2\ii R\bigr)^b}\,
\Biggl[\frac{4x_1x_2}
{(v_\mathrm{c}\tau-\ii r)(v_\mathrm{c}\tau+\ii r)}\Biggr]^c,
\label{eq:GRLcharge}\\
g_\mathrm{s}(\tau,x_1,x_2)&=&Z_1\,e^{\ii\frac{\pi}{4}}\Biggl[
\frac{1}{\pi}\,K_0\bigl(\Delta\sqrt{\tau^2+r^2/v_\mathrm{s}^2}\bigr)
+\int_{-\infty}^{\infty}\frac{d\theta}{2\pi}\,
K\bigl(\theta+\ii\tfrac{\pi}{2}\bigr)\,e^{\theta/2}\,
e^{2\ii\tfrac{\Delta}{v_\mathrm{s}}R\sinh\theta}\,
e^{-\Delta\tau\cosh\theta}+\ldots\Biggr],
\label{eq:GRLspin}
\end{eqnarray}
where $g_\mathrm{c,s}$ are the contributions of the charge and spin sectors
respectively.  Here $K_0$ is a modified Bessel function and the center-of-mass
coordinates are $R=(x_1+x_2)/2<0$ and $r=x_1-x_2<0$.  The normalization
constant $Z_1$ was obtained in Ref.~[\onlinecite{LukyanovZamolodchikov01}].
At the LEP~\cite{Ameduri-95} and the SU(2) invariant point 
the boundary reflection amplitude $K(\theta)$ is given by
\begin{equation}
\label{eq:defK}
K(\theta)=\ii\,\tanh\frac{\theta}{2}\quad \text{for}\ 
K_\mathrm{s}=\frac{1}{\sqrt{2}},
\quad
K(\theta)=
-\frac{\theta}{\pi^{3/2}}\,
\frac{\Gamma\bigl(\tfrac{\ii\theta}{\pi}\bigr)\,
\Gamma\bigl(\tfrac{3}{4}-\tfrac{\ii\theta}{2\pi}\bigr)}
{\Gamma\bigl(\tfrac{5}{4}+\tfrac{\ii\theta}{2\pi}\bigr)}\,
2^{-\frac{\ii}{\pi}\theta}\,\sinh\frac{\theta}{2}
\quad \text{for}\ K_\mathrm{s}=1.
\end{equation}
The expressions for general values of $K_\mathrm{s}$ can be found in
Refs.~[\onlinecite{GhoshalZamolodchikov94,MattssonDorey00,Caux-03}]. The
exponents in the charge sector are related to the Luttinger parameter by
\begin{equation}
a=\frac{1}{8}\left(K_\mathrm{c}+\frac{1}{K_\mathrm{c}}\right)^2,\quad 
b=\frac{1}{8}\left(K_\mathrm{c}-\frac{1}{K_\mathrm{c}}\right)^2,\quad
c=\frac{1}{8}\left(\frac{1}{K_\mathrm{c}^2}-K_\mathrm{c}^2\right).
\label{eq:exponents}
\end{equation}  
We stress that \eqref{eq:GRLspin} is independent of $\sigma$ and that
the dependence on $K_\mathrm{s}$ is through the overall normalization
constant $Z_1$ and the boundary reflection amplitude only.  The
one-particle contributions of the form-factor expansion are given by
the first two terms in \eqref{eq:GRLspin}, while the dots represent
corrections involving a higher number of particles in the intermediate
state as well as higher-order corrections due to the boundary. We have
determined the subleading terms in the spin part of the Green
function at the LEP and found their contribution to the LDOS
calculated below to be negligible (see
Sec.~\ref{sec:LDOShigherorders}).

After analytic continuation to real times $\tau\rightarrow \ii t$ the Green
function \eqref{eq:GRL} exhibits a light cone effect. The first term in
\eqref{eq:GRLspin} shows oscillating behavior for $r^2<(v_\mathrm{s}t)^2$ but
is damped otherwise. Similarly, the second term is oscillating for
$4R^2<(v_\mathrm{s}t)^2$. These oscillations are due to the propagation of
spinons from $x_2$ to $x_1$ either directly or via the boundary. In
particular, at late enough times both terms will oscillate. A similar light
cone effect was observed~\cite{SE07} in the Ising model with a boundary. On
the other hand, the charge part \eqref{eq:GRLcharge} possesses singularities
at $v_\mathrm{c}t=\pm r$ and $v_\mathrm{c}t=\pm 2R$ due to the propagation of
antiholons.

The small momentum regime of the Fourier transform of the LDOS is
obtained from ($\tau>0$, $x_1<x_2$)
\begin{eqnarray}
G^{RR}_{\sigma\sigma'}(\tau,x_1,x_2)&=&-\frac{\delta_{\sigma\sigma'}}{2\pi}
\frac{1}{\bigl(v_\mathrm{c}\tau-\ii r\bigr)^a}\,
\frac{1}{\bigl(v_\mathrm{c}\tau+\ii r\bigr)^b}\,
\Biggl[\frac{4x_1x_2}
{(v_\mathrm{c}\tau-2\ii R)(v_\mathrm{c}\tau+2\ii R)}\Biggr]^c\nonumber\\*
& &\hspace{-10mm}\times Z_1\Biggl[
\frac{1}{\pi}\,\sqrt{\frac{\ii\tau-r/v_\mathrm{s}}{\ii\tau+r/v_\mathrm{s}}}
\,K_{1/2}\bigl(\Delta\sqrt{\tau^2+r^2/v_\mathrm{s}^2}\bigr)
+\int_{-\infty}^{\infty}\frac{d\theta}{2\pi}\,
K\bigl(\theta+\ii\tfrac{\pi}{2}\bigr)\,
e^{2\ii\tfrac{\Delta}{v_\mathrm{s}}R\sinh\theta}\,
e^{-\Delta\tau\cosh\theta}+\ldots\Biggr]\label{eq:GRR}\\*
&=&G^{LL}_{\sigma\sigma'}(\tau,x_2,x_1).\nonumber
\end{eqnarray}
Compared to $G^{RL}$ the singularity at $v_\mathrm{c}t=2R$ is much softer
whereas the one at $v_\mathrm{c}t=r$ is more pronounced. 

\section{Local density of states}
\label{sec:LDOS}
The knowledge of the Green function \eqref{eq:GF} enables us to calculate
the LDOS, which is directly related to the tunneling current measured in STM
experiments. As was noted in Ref.~[\onlinecite{Kivelson-03}] it is useful to
consider the Fourier transform of the LDOS as physical properties can be more
easily identified. For example, this technique was used to study quasiparticle
interference in high-temperature
superconductors~\cite{Hoffman-02,Kohsaka-07}
and rare-earth compounds~\cite{Fang-07}.  We will first consider the boundary
condition $\Psi_\sigma(0)=0$, which results in the Green function
\eqref{eq:GRL} and hence yields a spin-independent LDOS.  More general
boundary conditions may lead to a spin-dependent LDOS or even the formation of
a boundary bound state. We will discuss this case the next section.

The Fourier transform of the LDOS is given by
$N_\sigma(E,Q)=N^>_\sigma(E,Q)+N^<_\sigma(E,Q)$, where
\begin{eqnarray}
N^>_\sigma(E,Q)&=&-\frac{1}{2\pi}
\int_{-\infty}^0 dx\int^\infty_{-\infty} dt\,e^{\ii(E t-Qx)}\,
G_{\sigma\sigma}(\tau>0,x,x)\Big|_{\tau\rightarrow\ii t+\delta},
\label{eq:defNp}\\*
N^<_\sigma(E,Q)&=&\frac{1}{2\pi}
\int_{-\infty}^0 dx\int^\infty_{-\infty} dt\,e^{\ii(E t-Qx)}\,
G_{\sigma\sigma}(\tau<0,x,x)\Big|_{\tau\rightarrow\ii t-\delta}.
\label{eq:defNm}
\end{eqnarray}
Here the Green function has been analytically continued to real
times and we have take the limit $x_1\rightarrow x_2\equiv x$. We will
focus on the LDOS for positive energies in what follows but note that
the LDOS for negative energies can be analyzed analogously. As
mentioned before, we will be mainly concerned with the
$2k_\mathrm{F}$-component as it vanishes in the absence of the
boundary and hence offers a particularly clean way of investigating
boundary effects.  For $Q\approx 2k_\mathrm{F}$ only
$G^{RL}_{\sigma\sigma}$ contributes and starting from \eqref{eq:GRL}
we arrive at the following expression (see App.~\ref{sec:FT})
\begin{eqnarray}
N^>_\sigma(E,2k_\mathrm{F}+q)&=&
\sum_{i=1}^2 N^>_i(E,2k_\mathrm{F}+q)+\ldots,\label{eq:Npos}\\
N^>_i(E,2k_\mathrm{F}+q)&=&-\Theta(E-\Delta)\,
\frac{Z_1\,e^{-\ii\frac{\pi}{4}(4c+1)}\,\Gamma(2c+1)}
{8\pi^2\,v_\mathrm{c}^{a+b-1}\,\Gamma(a+b+2c)}\nonumber\\*
& &\qquad\times\int_{-A}^{A} d\theta 
\frac{h_i(\theta)\,u_i^{2c+1}}{(E\!-\!\Delta\cosh\theta)^{2-a-b}}
\,F_1\bigl(2c+1,a,b,a+b+2c;u_i^*,-u_i\bigr).
\label{eq:Ni}
\end{eqnarray}
Here $|q|\ll 2k_\mathrm{F}$, $A=\mathrm{arcosh}\bigl(\tfrac{E}{\Delta}\bigr)$,
$F_1$ denotes Appell's hypergeometric function~\cite{ErdelyiHTF1} (see
App.~\ref{sec:hf}), $h_1(\theta)=1$,
$h_2(\theta)=K\bigl(\theta+\ii\tfrac{\pi}{2}\bigr)\,e^{\theta/2}$, and
\begin{equation}
u_1=\frac{2}{v_\mathrm{c}q}\bigl(E-\Delta\cosh\theta\bigr)+
\ii\,\text{sgn}\Bigl(\frac{v_\mathrm{s}q}{\Delta}\Bigr)\delta, \quad
u_2=\frac{2v_\mathrm{s}}{v_\mathrm{c}}\,
\frac{E-\Delta\cosh\theta}{v_\mathrm{s}q-2\Delta\sinh\theta}+
\ii\,\text{sgn}\Bigl(\frac{v_\mathrm{s}q}{\Delta}-2\sinh\theta\Bigr)\delta,
\label{eq:ui}
\end{equation}
where $\delta\rightarrow 0+$. The result \eqref{eq:Ni} is valid for $a+b<2$
and $-1/2<c$. Below we plot $N^>_\sigma(E,2k_\mathrm{F}+q)$ for two different
parameter regimes. We smear out singularities by taking $\delta$ small but
finite ($\delta=0.01$ unless stated), which mimics broadening by instrumental
resolution and temperature in experiments. The results presented below apply
to the regime $T\ll E,\Delta,v_\mathrm{c}/a_0$ ($a_0$ is the lattice spacing),
where temperature effects are negligible.

\subsection{Repulsive Case}
\begin{figure}[t]
\centering
\includegraphics[scale=0.34,clip=true]{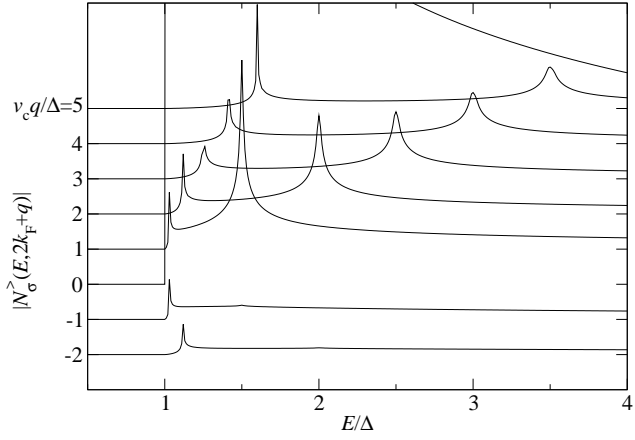}
\caption{$|N^>_\sigma(E,2k_\mathrm{F}+q)|$ (arbitrary units) for 
  $K_\mathrm{c}=0.8$, $K_\mathrm{s}=1$ and $v_\mathrm{c}=2v_\mathrm{s}$. The
  curves are constant $q$-scans which have been offset along the y-axis by a
  constant with respect to one another.  The LDOS is dominated by a strong
  peak at $q=0$, i.e. $Q=2k_\mathrm{F}$. We further observe dispersing features
  at $E_\mathrm{c}=v_\mathrm{c}|q|/2+\Delta$ and
  $E_\mathrm{s}=\sqrt{(v_\mathrm{s}q/2)^2+\Delta^2}$.  For $q<0$ the
  dispersing features are strongly suppressed.}
\label{fig:plot1}
\end{figure}
We first consider the case $v_\mathrm{s}<v_\mathrm{c}$,
$K_\mathrm{c}<1$. This can be thought of as providing a simplified
model for the LDOS of a two-leg ladder with repulsive
electron-electron interactions. The low-energy theory for the ladder
is similar in that there is a gapless charge sector and a gapped spin
sector, but the full description of the latter is considerably more
complicated\cite{soN}.

\begin{figure}[t]
\centering
\includegraphics[scale=0.34,clip=true]{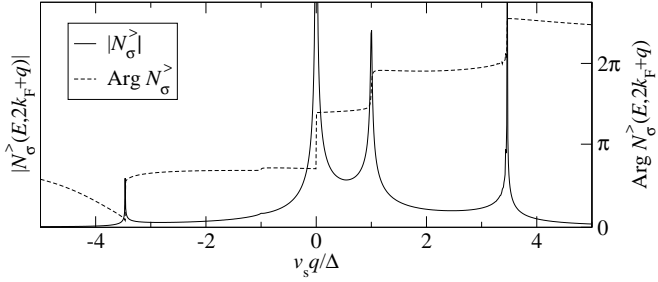}
\caption{Constant energy scan for $E=2\Delta$: 
  $|N^>_\sigma(E,2k_\mathrm{F}+q)|$ (arbitrary units) and
  $\text{Arg}N^>_\sigma(E,2k_\mathrm{F}+q)$ for $K_\mathrm{c}=0.8$,
  $K_\mathrm{s}=1$ and $v_\mathrm{c}=2v_\mathrm{s}$. We observe a peak at
  $q=0$ (related to the pinning of the CDW at the boundary) and dispersing
  features at $q=\pm 2(E-\Delta)/v_\mathrm{c}$ as well as $q=\pm
  2\sqrt{E^2-\Delta^2}/v_\mathrm{s}$. For $q<0$ the dispersing features are
  strongly suppressed. Furthermore, we observe characteristic jumps in the
  argument at the positions of the peaks.}
\label{fig:plot2}
\end{figure}

In Figs.~\ref{fig:plot1} and~\ref{fig:plot2} we plot
$N^>_\sigma(E,2k_\mathrm{F}+q)$ for the case of unbroken spin rotational
symmetry ($K_\mathrm{s}=1$).  The Fourier transform of the LDOS is dominated
by a singularity at momentum $2k_\mathrm{F}$ ($q=0$), which arises from the
contribution $N^>_1$. For fixed energy and close to the singularity this term
behaves as (see App.~\ref{sec:singularities})
\begin{equation}
N^>_1(E,2k_\mathrm{F}+q)\sim 
\left(\frac{1}{v_\mathrm{c}q}\right)^\alpha,
\quad \alpha=1-\frac{K_\mathrm{c}^2}{2},
\label{eq:CDWpeak}
\end{equation}
which implies a phase jump of $\pi\alpha$ as $q\rightarrow 0\pm$.  This peak
is indicative of the CDW order being pinned at the boundary. We note, however,
that the peak occurs at finite energies and hence the underlying
process is not static.  A similar feature can be seen in 
the Luttinger liquid case~\cite{Kivelson-03}, where the singularity as a
function of $q$ is softer ($\alpha_\mathrm{LL}=(1-K_\mathrm{c}^2)/2$). 

At low energies above the spin gap $\Delta$ we
further observe two dispersing features associated with the collective spin
and charge degrees of freedom respectively. These are broadly similar
to the \emph{bulk} single-particle spectral function
\cite{bulkspectralfunction,Mottbulk} and feature (1) a ``charge peak''
that follows 
\begin{equation}
E_\mathrm{c}(q)=\frac{v_\mathrm{c}|q|}{2}+\Delta
\label{eq:cdispersion}
\end{equation} 
and (2) a ``spin peak'' at position
\begin{equation}
E_\mathrm{s}(q)=\sqrt{\left(\frac{v_\mathrm{s}q}{2}\right)^2+\Delta^2}.
\label{eq:sdispersion}
\end{equation} 
We note that neither peak is sharp (i.e. they are not delta-functions) and
hence have to be thought of as arising from excitations involving at least two
``elementary'' constituents.  In this way of thinking, the charge peak arises
from two-particle excitations composed of a ``zero momentum'' spinon
contributing an energy $\Delta$ and a gapless antiholon of ``momentum'' $q$.
On the other hand, the spin peak can be thought of arising from two-particle
excitations composed of a ``zero momentum'' antiholon and a spinon of
``momentum'' $q$. The appearance of $v_\mathrm{c}/2$ and $v_\mathrm{s}/2$ in
\eqref{eq:cdispersion} and \eqref{eq:sdispersion}, respectively, is due to the
fact that the particles have to propagate to the boundary and back, thus
covering the distance $2x$ in time $t$.  We note that on a technical level the
charge peak arises from the contribution $N^>_1$ to the Fourier transform of
the LDOS, whereas the spin peak has its origin in $N^>_2$, which encodes the
effects of the boundary on the spin degrees of freedom. In the $q<0$ region
the dispersing features are strongly suppressed and for $K_\mathrm{c}=1$ the
charge feature is found to vanish entirely.

It is instructive to plot $N^>_\sigma(E,2k_\mathrm{F}+q)$ as a
function of $q$ for fixed energy, see Fig.~\ref{fig:plot2}.
We observe characteristic jumps in the phase $\text{Arg}\,N^>_\sigma$
at the peak positions. This is similar to the Luttinger liquid
case~\cite{Kivelson-03}.  

\subsection{Attractive case}\label{sec:LDOSattractive}
We now turn to the case of a CDW state arising in a system with (effective)
attractive electron-electron interactions. As discussed in
Sec.~\ref{sec:model}, this case arises in electron-phonon systems. The
effective parameters are given by $v_\mathrm{s}>v_\mathrm{c}$ and
$K_\mathrm{c}>1>K_\mathrm{s}$.

\begin{figure}[t]
\centering
\includegraphics[scale=0.34,clip=true]{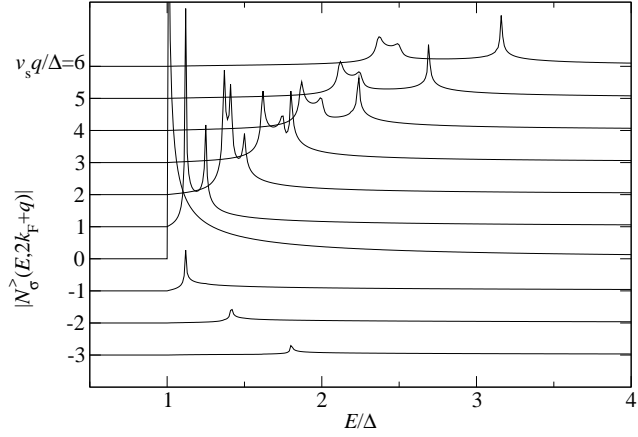}
\caption{$|N^>_\sigma(E,2k_\mathrm{F}+q)|$ (arbitrary units) for 
  $K_\mathrm{c}=1.2$, $K_\mathrm{s}=1$ and $v_\mathrm{s}=2v_\mathrm{c}$.  The
  curves are constant $q$-scans which have been offset along the y-axis by a
  constant with respect to one another. The peak at $q=0$ is much less
  pronounced then in the repulsive case (see Fig.~\ref{fig:plot1}). We observe
  dispersing features at $E_\mathrm{c}$, $E_\mathrm{s}$, and
  $E_\mathrm{cs}=v_\mathrm{c}|q|/2+
  \Delta\sqrt{1-(v_\mathrm{c}/v_\mathrm{s})^2}$ (for $|q|>q_0$ only).}
\label{fig:plot3}
\end{figure}

\begin{figure}[t]
\centering
\includegraphics[scale=0.34,clip=true]{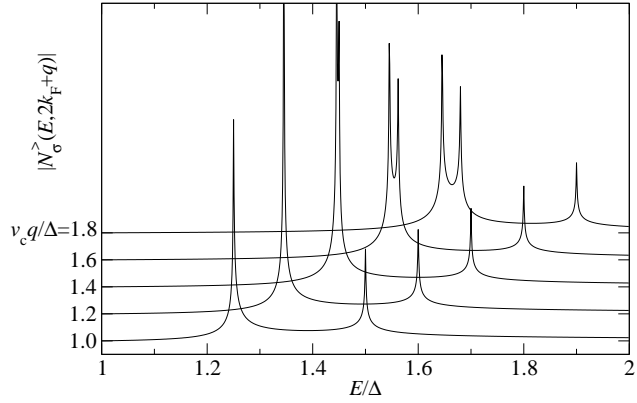}
\caption{$|N^>_\sigma(E,2k_\mathrm{F}+q)|$ (arbitrary units) for 
  $K_\mathrm{c}=1$, $K_\mathrm{s}=1/\sqrt{2}$,
  $v_\mathrm{s}=1.5\,v_\mathrm{c}$, $\delta=0.001$, and
  $v_\mathrm{c}q/\Delta=1.0,\ldots,1.8$.  The curves have been offset along
  the y-axis by a constant with respect to one another. We observe the
  splitting of the spin peak at $E_\mathrm{s}$ at the critical momentum
  $v_\mathrm{c}q_0/\Delta\approx 1.19$.}
\label{fig:plot4}
\end{figure}

In Fig.~\ref{fig:plot3} we plot $N^>_\sigma(E,2k_\mathrm{F}+q)$ as a
function of energy for several values of $q$ (in units of the spin gap).
We again observe a singularity at $2k_\mathrm{F}$, which arises from
\eqref{eq:CDWpeak}. The singularity is much less pronounced than in
the repulsive case and disappears for $K_\mathrm{c}\ge\sqrt{2}$. 
Like in the repulsive case there are several dispersing features:
\begin{enumerate}
\item{} a charge peak at $E=E_\mathrm{c}(q)$, where $E_\mathrm{c}$ is given by
(\ref{eq:cdispersion}); 

\item{} a spin peak at $E=E_\mathrm{s}(q)$, where $E_\mathrm{s}$ is defined in
(\ref{eq:sdispersion}); 

\item{} when $|q|$ exceeds a critical value $q_0$ a third dispersing
  low-energy peak appears (see Fig.~\ref{fig:plot4}) at
\begin{equation}
E_\mathrm{cs}(q)=\frac{v_\mathrm{c}|q|}{2}+
\Delta\sqrt{1-\left(\frac{v_\mathrm{c}}{v_\mathrm{s}}\right)^2}
=E_\mathrm{s}(q_0)+\frac{v_\mathrm{c}}{2}\bigl(|q|-q_0\bigr),\quad
q_0=\frac{2\Delta v_\mathrm{c}}{v_\mathrm{s}
\sqrt{v_\mathrm{s}^2-v_\mathrm{c}^2}}.
\label{eq:scdispersion}
\end{equation}  
This feature can be thought of as arising from a ``momentum'' $q_0$
spinon and an antiholon carrying ``momentum'' $q-q_0$. We 
note that in this case the spin and charge excitations have the same
group velocity 
\begin{equation}
\frac{\partial E_\mathrm{c}}{\partial q}=
\frac{\partial E_\mathrm{s}}{\partial q}\bigg|_{q=q_0}=
\frac{v_\mathrm{c}}{2}.
\end{equation}
\end{enumerate}
This behavior is reminiscent of what is found for the single-particle
spectral function in the
bulk~\cite{bulkspectralfunction,Mottbulk}. The peak splitting and
hence the qualitative difference between the repulsive and attractive
regime is a consequence of the curvature of the (anti)soliton
dispersion relation and hence of the spin gap.  In the Luttinger
liquid case~\cite{Kivelson-03} (where both sectors are massless) there
are only two dispersing features in both regimes.

\subsection{Higher-order corrections}\label{sec:LDOShigherorders}
As we have indicated in Eq.~(\ref{eq:GRLspin}) there are contributions
to the LDOS beyond those that we have discussed above. They arise from
our calculation of the spin-part of the Green function and are
expected to be small\cite{SE07}. In order to verify that they can
indeed be neglected, we have analyzed them in some detail at the
LEP $K_\mathrm{s}=1/\sqrt{2}$, where the necessary
matrix elements take a particularly simple form, which makes the
actual calculations much easier.

Our purpose is then to determine further terms in the the expansion
\eqref{eq:Npos} of $N^>_\sigma(E,2k_\mathrm{F}+q)$.  We denote by
$N_{nm}$ the contribution to \eqref{eq:Npos} that arises from
processes in which $n$ gapped spinons (which correspond to
solitons/antisolitons in the sine-Gordon model describing the spin
sector) propagate between $(0,x_2)$ and $(\tau,x_1)$, and which
involves the $m$'th power of the boundary reflection matrix $K$.  The
terms discussed above correspond to $N_{10}=N^>_1$ and $N_{01}=N^>_2$.
In App.~\ref{sec:appho} we present some details on the calculation of
the terms in the form-factor expansion of $g_s(\tau,x_1,x_2)$ that
give rise to $N_{nm}$ with $m+n\le 3$. The number of
$\theta$-integrations in $N_{nm}$ equals $m+n$ (cf. \eqref{eq:Ni}).
We find that their contributions to the
Fourier transform of the LDOS are small. In particular, all
qualitative features of the LDOS such as dispersing peaks are already
encoded in $N_{10}$ and $N_{01}$.
\begin{figure}[t]
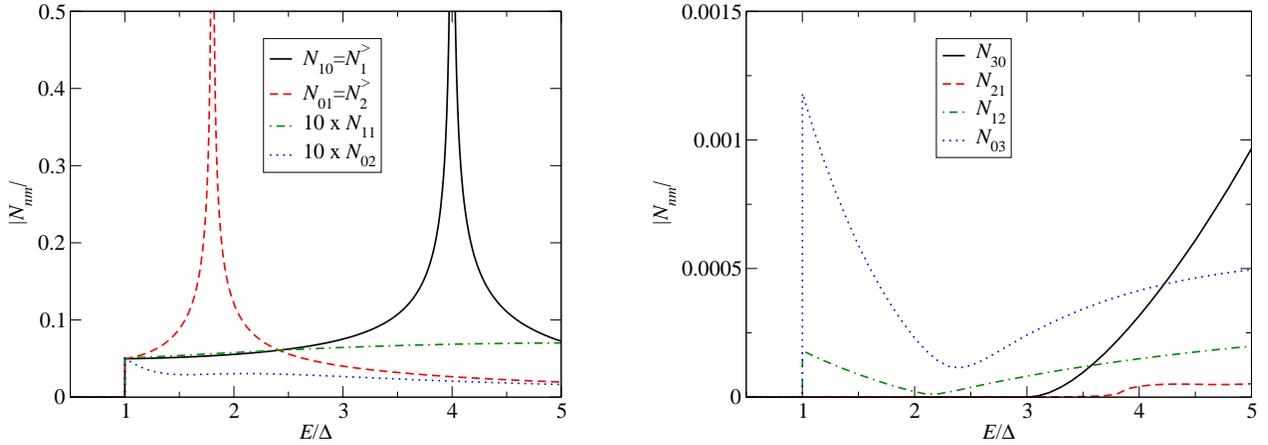

\centering
\includegraphics[scale=0.34,clip=true]{plot5a.eps}\hspace{10mm}
\includegraphics[scale=0.34,clip=true]{plot5b.eps}
\caption{(Color online) Left: Comparison of the absolute values of
  $N_{10}=N^>_{1}(E,2k_\mathrm{F}+q)$ and
  $N_{01}=N^>_{2}(E,2k_\mathrm{F}+q)$ to the sub-leading terms
  $N_{11}$ and $N_{02}$ ($N_{20}=0$). We stress that the scale for the
  higher-order terms has been magnified. Right: Absolute values of the
  terms $N_{30}$, $N_{21}$, $N_{12}$, and $N_{03}$. We stress the
  different scales on the y-axis. The parameters are 
  $K_\mathrm{c}=1$, $K_\mathrm{s}=1/\sqrt{2}$,
  $v_\mathrm{c}=2v_\mathrm{s}$, and $v_\mathrm{s}q/\Delta=3$. The
  three-particle contributions $N_{30}$ and $N_{21}$ vanish for
  $E<3\Delta$. Furthermore, the higher-order terms possess no peaks.}
\label{fig:plot5}
\end{figure}
In Fig.~\ref{fig:plot5} we show the leading terms $N_{10}$ and
$N_{01}$ as well as the sub-leading terms 
for $K_\mathrm{c}=1$ and $v_\mathrm{c}>v_\mathrm{s}$ for a fixed value of $q$
as a function of energy.  We see that the two leading terms in \eqref{eq:Npos}
indeed capture all qualitative features of the LDOS and carry the main part of
the spectral weight at low energies $E\le 5\Delta$. The higher-order terms are
small compared to $N_{10}$ and $N_{01}$.  In particular, $N_{30}$ vanishes for
$E<3\Delta$, since this term originates from a three-particle process. In
general, all terms $N_{nm}$ originating from $n$-particle processes vanish for
$E<n\Delta$. Most importantly, however, the higher-order terms do not possess
any singularities.  The suppression of subleading terms in the form-factor
expansion for bulk two-point functions is a well-known feature of massive
theories~\cite{YurovZamolodchikov91,CardyMussardo93,EsslerKonik05}, whereas
the smallness of terms involving higher powers of the boundary reflection
amplitude $K$ has recently been demonstrated for the Ising model with a
boundary magnetic field~\cite{SE07}.

\subsection{Small-momentum regime}\label{sec:LDOSsmr}
The small-momentum regime $Q\approx 0$ of the Fourier transform of the
LDOS can be analyzed in the same way as in the $Q\approx 2k_\mathrm{F}$ case
discussed above. We note that the LDOS for $Q\approx 0$ is non-vanishing
even in absence of a boundary~\cite{bulkspectralfunction,Mottbulk}. In the
presence of a boundary the Fourier transform of the LDOS for $Q\approx
0$ is obtained from \eqref{eq:GRR}. The leading terms are given by
$N^>_\sigma(E,Q)= \sum_{i=1}^2 N^>_i(E,Q)+\ldots$, where
\begin{eqnarray}
N^>_i(E,Q)&=&-\Theta(E-\Delta)\,
\frac{Z_1\,e^{-\ii\frac{\pi}{2}(2c+1)}\,\Gamma(a+b+1)}
{4\pi^2\,v_\mathrm{c}^{a+b-1}\,\Gamma(2a+2b)}\nonumber\\*
&&\qquad\times\int_{-A}^{A} d\theta
\frac{h_i(\theta)\,u_i^{2c+1}}{(E\!-\!\Delta\cosh\theta)^{2-a-b}}
\,F_1\bigl(a+b+1,c,c,2a+2b;u_i^*,-u_i\bigr).
\label{eq:NiRR}
\end{eqnarray}
Here we have $|Q|\ll k_\mathrm{F}$,
$A=\mathrm{arcosh}\bigl(\tfrac{E}{\Delta}\bigr)$, $h_1(\theta)=e^{\theta/2}$,
$h_2(\theta)=K\bigl(\theta+\ii\tfrac{\pi}{2}\bigr)$, and $u_{1,2}$ are defined
in \eqref{eq:ui} with $q$ replaced by $Q$. The main difference to
\eqref{eq:Npos} is in the dependence of Appell's hypergeometric function on
$K_\mathrm{c}$.

\begin{figure}[t]
\centering
\includegraphics[scale=0.34,clip=true]{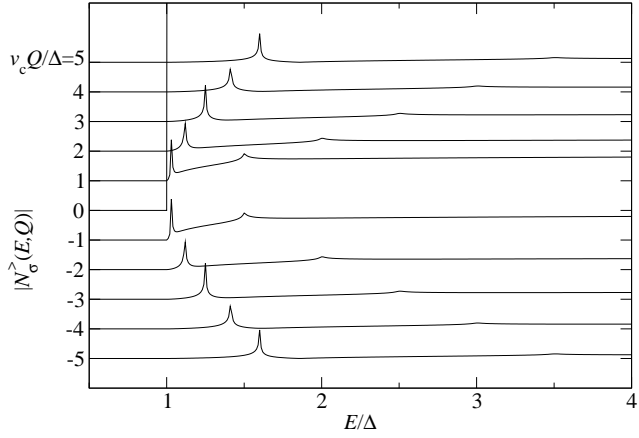}
\caption{$|N^>_\sigma(E,Q)|$ (arbitrary units) for 
  $K_\mathrm{c}=0.8$, $K_\mathrm{s}=1$ and $v_\mathrm{c}=2v_\mathrm{s}$.  The
  curves are constant $Q$-scans which have been offset along the y-axis by a
  constant with respect to one another. We observe dispersing features at
  $E_\mathrm{c}(Q)$ and $E_\mathrm{s}(Q)$. The charge features are
  very weak for all momenta. We further observe a constant background
  at energies larger than the spin gap.}
\label{fig:plot6}
\end{figure}
\begin{figure}[t]
\centering
\includegraphics[scale=0.34,clip=true]{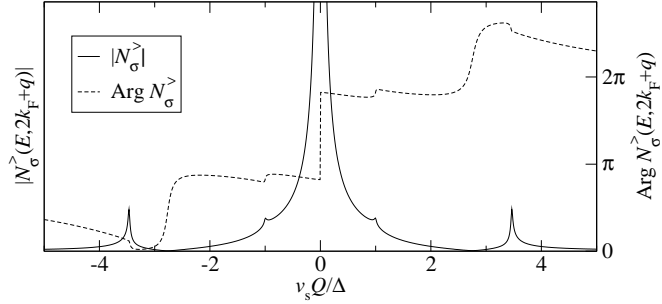}
\caption{Constant energy scan for $E=2\Delta$: 
  $|N^>_\sigma(E,Q)|$ (arbitrary units) and
  $\text{Arg}N^>_\sigma(E,Q)$ for $K_\mathrm{c}=0.8$,
  $K_\mathrm{s}=1$ and $v_\mathrm{c}=2v_\mathrm{s}$. We observe a strong peak
  at $Q=0$ and dispersing features at $Q=\pm 2(E-\Delta)/v_\mathrm{c}$ as well
  as $Q=\pm 2\sqrt{E^2-\Delta^2}/v_\mathrm{s}$. Note that $|N^>_\sigma(E,Q)|$
  is symmetric under $Q\rightarrow-Q$.}
\label{fig:plot7}
\end{figure}

In Figs.~\ref{fig:plot6} and~\ref{fig:plot7} we plot $N^>_\sigma(E,Q)$
for the case of repulsive electron interactions and unbroken spin
rotational symmetry.  It is dominated by a singularity at $Q=0$, which
has its origin in $N_1^>$ and behaves as $\sim 1/Q$ independently of
$K_\mathrm{c}$.  This singularity is more pronounced than its
counterpart at $2k_\mathrm{F}$.  We further observe dispersing
features at positions $E_\mathrm{c}(Q)$ and $E_\mathrm{s}(Q)$
respectively. Both of these are symmetric under $Q\rightarrow -Q$.
The peak at $E_{\mathrm{c}}(Q)$ is strongly suppressed, vanishes for
$K_\mathrm{c}=1$, and becomes a dip in the attractive regime.  The
suppression is due to the softness of the singularities of $G^{RR}$ at
$v_\mathrm{c}t=2R$. On the other hand, the charge part of
\eqref{eq:GRR} has its strongest singularity at $v_\mathrm{c}t=r=0$,
which results in a background of spectral weight in $N^>_\sigma(E,Q)$
for all energies above the spin gap.

In the attractive case ($v_\mathrm{s}>v_\mathrm{c}$) we observe a
peak-splitting similiar to that in the $2k_\mathrm{F}$-component, but all peaks
are very weak.

\section{General boundary conditions and boundary bound states}
\label{sec:generalBC}
So far we have considered the simplest possible boundary conditions
corresponding to a spin-independent phase shift of $\pi$. Both ways of
realizing a boundary in a (quasi) one dimension system that we have
discussed above (i.e. as a result of an impurity or in a ``two-tip''
STS experiment) are expected to give rise to a local potential or
magnetic field. These correspond to more general phase shifts for
reflection of particles at the boundary. As is well known, such
more general boundary conditions can give rise to boundary bound
states, see for example
Refs.~[\onlinecite{KapustinSkorik96}]. These are
expected to be visible in the Fourier transform of the LDOS as
``resonances'' inside the single-particle gap. This is most easily
seen by considering a Lehmann representation of $N^>_\sigma(E,Q)$ in
terms of the eigenstates $|n_\mathrm{b}\rangle$ on the half-line
\begin{equation}
  N^>_\sigma(E,Q)=\int_{-\infty}^0 dx\,e^{-\ii Qx}\,\sum_{n_\mathrm{b}}\,
  \big|\!\bra{0_\mathrm{b}}\Psi_\sigma(0,x)\ket{n_\mathrm{b}}\big|^2\,
  \delta(E-E_{n_\mathrm{b}}).
\end{equation}
For boundary bound states $|{\rm bbs},\alpha\rangle$ we have $0<E_{{\rm
    bbs},\alpha}<\Delta$, which leads to features in $N^>_\sigma(E,Q)$ below
the single-particle gap. As we are dealing with a spin-charge separated
system, these features will generally not be sharp as the bound state occurs
only in the gapped sector of the theory. We now turn to calculating the LDOS
in cases where boundary bound states exist.  We first consider boundary
conditions of the form 
\be 
R_\sigma(\tau,0)=-e^{-\ii f_\sigma\Phi_\mathrm{s}^0/2}\, L_\sigma(\tau,0), 
\ee 
where $f_\up=1=-f_\dw$.  In terms of the Bose fields these boundary conditions
read 
\be 
\Phi_\mathrm{c}(\tau,0)=0,\qquad 
\Phi_\mathrm{s}(\tau,0)=\Phi_\mathrm{s}^0,\quad 
0\le\Phi_\mathrm{s}^0<\pi.
\label{newbcs}
\ee 
We note that these boundary conditions break spin rotational symmetry.
However, if we go over to the case of a one-dimensional Mott insulator by
exchanging spin and charge degrees of freedom, the spin rotational symmetry
remains intact and the boundary conditions correspond to a local potential.

As before we will focus on the $2k_\mathrm{F}$-component of the
Fourier transform of the LDOS. As we have changed the boundary
conditions only in the spin sector, the charge part \eqref{eq:GRLcharge} of
the chiral Green function remains unchanged. 

The two leading terms of the form-factor expansion in the spin sector
are still of the form \eqref{eq:GRLspin}, but now the boundary
reflection amplitude $K$ is different and in particular is
spin-dependent. At the LEP it is given
by~\cite{Ameduri-95}
\begin{equation}
K^{\sigma\bar{\sigma}}(\theta)=
\frac{\sin\Bigl(\ii\frac{\theta}{2}-
f_\sigma\frac{\Phi_\mathrm{s}^0}{2}\Bigr)}
{\cos\Bigl(\ii\frac{\theta}{2}+
f_\sigma\frac{\Phi_\mathrm{s}^0}{2}\Bigr)},\quad
K^{\sigma\sigma}(\theta)=K^{\bar{\sigma}\bar{\sigma}}(\theta)=0.
\label{eq:KLE}
\end{equation}
Here we have introduced the notations $\up=+$ and $\dw=-$ as well as
$\bar{\sigma}=-$ for $\sigma=+$ and vice versa. We note that as a result of a
different choice of phase for the asymptotic states Eq.~(\ref{eq:KLE}) differs
by a minus sign from Ref.~[\onlinecite{Ameduri-95}] (see also
App.~\ref{sec:appgfss}). The expressions for general $K_\mathrm{s}$ can be
found in Refs.~[\onlinecite{GhoshalZamolodchikov94,MattssonDorey00,Caux-03}].

In the spin rotationally symmetric case $K_\mathrm{s}=1$ we have
$K^{+-}(\theta)=K^{-+}(\theta)=K(\theta)$, where $K(\theta)$ is given in
\eqref{eq:defK}.  We stress that the Green function remains diagonal in
spin space, $G^{RL}_{\sigma\sigma'}\propto\delta_{\sigma\sigma'}$, and that
the spin-dependence is entirely due to the boundary reflection matrix
$K^{\sigma\bar{\sigma}}$. Before presenting the resulting LDOS we discuss
the emergence of a boundary bound state in the spin
sector~\cite{GhoshalZamolodchikov94,SkorikSaleur95,MattssonDorey00}. If
we choose the phase-shift $\Phi_\mathrm{s}^0$ in the spin sector such that
\be
K_\mathrm{s}^2 \pi < \Phi_\mathrm{s}^0,
\ee
the boundary reflection amplitude $K^{-+}(\theta)$ has a pole in the
physical strip $0\le\mathfrak{Im}\,\theta\le\pi/2$. This pole
corresponds to a boundary bound state with energy
\begin{equation}
E_\mathrm{bbs}=\Delta\,\sin\gamma,
\quad \gamma=\frac{\pi-\Phi_\mathrm{s}^0}{2-2K_\mathrm{s}^2}.
\label{eq:bbsenergy}
\end{equation}
The physical nature of the bound state has been discussed by Ghoshal
and Zamolodchikov~\cite{GhoshalZamolodchikov94}.
The classical ground state of the sine-Gordon model on the half-line
\eqref{eq:spinhamiltonian} in the entire range $0\le\Phi_\mathrm{s}^0<\pi$ 
is characterized by the asymptotic behavior
$\Phi_\mathrm{s}\rightarrow 0$ as $x\rightarrow -\infty$. 
On the other hand, there exists a second classically stable state
satisfying $\Phi_\mathrm{s}\rightarrow 2\pi$ as $x\rightarrow
-\infty$. When $\Phi_\mathrm{s}^0$ is sufficiently large this state
is expected to be stable in the quantum theory as well. 

We note that for $\Phi_\mathrm{s}^0=\pi$ both states are degenerate
and \eqref{eq:bbsenergy} vanishes. In the attractive regime of the
sine-Gordon model $K_\mathrm{s}<1/\sqrt{2}$ additional boundary bound
states occur, while in the spin rotationally invariant case
$K_\mathrm{s}\rightarrow 1$ the condition
$K_\mathrm{s}^2\pi<\Phi_\mathrm{s}^0<\pi$ is never satisfied and
hence no boundary bound states exist.

When calculating dynamical response functions in the boundary state
formalism additional contributions in the form factor expansions occur
upon analytical continuation in the rapidity variables.  In
particular, the pole of the boundary reflection amplitude in the
physical strip gives rise to an additional term linear in $K$ in the
form-factor expansion \eqref{eq:GRLspin}. In the case $\tau>0$ and
$x_1<x_2$ it takes the form (see App.~\ref{sec:appbbs})
\begin{equation}
\Theta\Bigl(\Phi_\mathrm{s}^0-K_\mathrm{s}^2\,\pi\Bigr)\,
\delta_{\sigma\dw}\,Z_1\,B\,e^{\frac{\ii}{2}\gamma}\,
e^{2\frac{\Delta}{v_\mathrm{s}}R\cos\gamma}\,
e^{-\Delta\tau\sin\gamma},
\label{eq:GRLbbs}
\end{equation}
where the constant $B\ge 0$ is related to the residue of $K^{\mp\pm}(\theta)$
(see (\ref{constantB})). At the LEP it equals
$B=-2\cos\Phi_\mathrm{s}^0$. We stress that this additional term
appears in the down-spin channel only, since we have assumed
$0\le\Phi_\mathrm{s}^0<\pi$. If we were to consider
$-\pi<\Phi_\mathrm{s}^0\le 0$, we would find a term similar to
\eqref{eq:GRLbbs} in the up-spin channel only.

The Fourier transform of the LDOS for the boundary conditions
(\ref{newbcs}) can be expanded as before and is expressed as
\begin{equation}
N^>_\sigma(E,2k_\mathrm{F}+q)=\sum_{i=1}^3
N^>_{\sigma,i}(E,2k_\mathrm{F}+q)+\ldots. 
\label{eq:Nposbbs}
\end{equation}
Here the first two terms are again of the form \eqref{eq:Ni}, 
where in the second term $N^>_{\sigma,2}$ we need to replace the
boundary reflection amplitude $K\bigl(\theta+\ii\frac{\pi}{2}\bigr)$
by its spin-dependent counterpart
$K^{\sigma\bar{\sigma}}\bigl(\theta+\ii\frac{\pi}{2}\bigr)$. The 
third term is obtained from \eqref{eq:GRLbbs} and arises as a result
of the presence of a boundary bound state. Explicitly it reads
\begin{eqnarray}
N^>_{\sigma,3}(E,2k_\mathrm{F}+q)&=&
\Theta\Bigl(\Phi_\mathrm{s}^0-K_\mathrm{s}^2\,\pi\Bigr)\,
\Theta\Bigl(E-E_\mathrm{bbs}\Bigr)\,\delta_{\sigma\dw}\,
\frac{Z_1B}{4\pi}\,\frac{\Gamma(2c+1)}{\Gamma(a+b+2c)}\,
\frac{\ii\,e^{\frac{\ii}{2}\gamma}\,e^{-\ii\pi c}}{v_\mathrm{c}^{a+b-1}}
\nonumber\\*
&&\times 
\frac{\left(\frac{2}{v_\mathrm{c}q}(E-E_\mathrm{bbs})
+\ii\,\sgn{\frac{v_\mathrm{s}q}{\Delta}}\delta\right)^{2c+1}}
{\bigl(E-E_\mathrm{bbs}\bigr)^{2-a-b}}\,
F_\mathrm{D}^{(3)}\bigl(2c+1,a,b,2c,a+b+2c;u_3^*,-u_3,-u_3'\bigr).
\label{eq:N3}
\end{eqnarray}
Here $F_\mathrm{D}^{(3)}$ denotes Lauricella's hypergeometric function of
three arguments~\cite{Lauricella93} (see App.~\ref{sec:hf}) and 
\begin{equation}
u_3=\frac{2}{v_\mathrm{c}q}(E-E_\mathrm{bbs})
+\ii\frac{2\Delta}{v_\mathrm{s}q}\cos\gamma,\quad
u_3'=\ii\frac{2\Delta}{v_\mathrm{s}q}\cos\gamma,
\end{equation}
where the constant $\gamma$ defined in \eqref{eq:bbsenergy}.  The Fourier
transform of the LDOS \eqref{eq:Nposbbs} has a non-dispersing singularity at
its lower threshold
\begin{equation}
N^>_{\sigma,3}(E,2k_\mathrm{F}+q)\propto 
\frac{\delta_{\sigma\dw}}{(E-E_\mathrm{bbs})^{\alpha}}, 
\quad \alpha=1-\frac{1}{2K_\mathrm{c}^2}.
\label{eq:Nbbssingularity}
\end{equation}
The emergence of a non-dispersing feature within the spin gap signals the
presence of a boundary bound state.  In Fourier space the LDOS is a
convolution of contributions from the spin and charge sectors. As
we are dealing with a bound state in the spin sector, the exponent of
the singularity depends only on the Luttinger parameter in the charge
sector. We note that the singularity occurs only in the down-spin
channel and disappears for $K_\mathrm{c}^2\le 1/2$. On the other hand,
in $N^<_\sigma$ the additional feature due to the boundary bound state
appears only in the up-spin channel.

\begin{figure}[t]
\centering
\includegraphics[scale=0.34,clip=true]{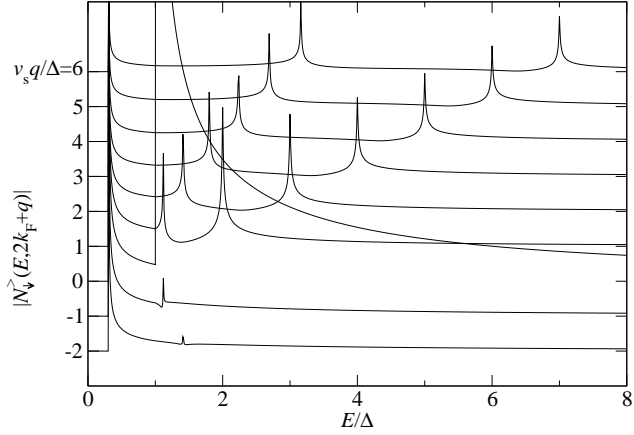}
\caption{$|N^>_\dw(E,2k_\mathrm{F}+q)|$ (arbitrary units) for 
  $K_\mathrm{c}=1$, $K_\mathrm{s}=1/\sqrt{2}$, $v_\mathrm{c}=2v_\mathrm{s}$,
  and $\Phi_\mathrm{s}^0=0.9\pi$. The curves are constant $q$-scans which have
  been offset along the y-axis by a constant with respect to one another.  We
  observe dispersing features at $E_\mathrm{s}(q)$ and
  $E_\mathrm{c}(q)$ (for $q>0$ only) as well as a non-dispersing
  singularity at $E=E_\mathrm{bbs}$, which is due to the formation of
  a boundary bound state in the spin sector.} 
\label{fig:plot8}
\end{figure}
In Fig.~\ref{fig:plot8} we plot the down-spin component of \eqref{eq:Nposbbs}
for $v_\mathrm{c}>v_\mathrm{s}$ as a function of energy for several
values of $q$.  As before, at low energies above the spin
gap $\Delta$ we observe two dispersing features associated with the collective
spin and charge degrees of freedom that follow $E_\mathrm{s}$ and
$E_\mathrm{c}$ respectively. In addition, we observe the non-dispersing
singularity \eqref{eq:Nbbssingularity} at $E=E_\mathrm{bbs}$. 

\begin{figure}[t]
\centering
\includegraphics[scale=0.34,clip=true]{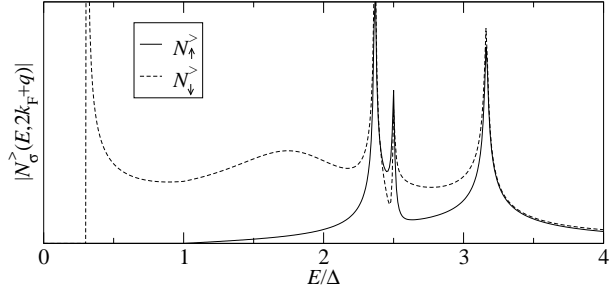}
\caption{$|N^>_\up(E,2k_\mathrm{F}+q)|$ (full line)
  and $|N^>_\dw(E,2k_\mathrm{F}+q)|$ (dashed line) for
  $v_\mathrm{s}q/\Delta=6$, $K_\mathrm{c}=1$, $K_\mathrm{s}=1/\sqrt{2}$,
  $v_\mathrm{s}=2v_\mathrm{c}$, and $\Phi_\mathrm{s}^0=0.9\,\pi$.  The broad
  maximum at $E\approx 1.8$ is caused by the excitation of the boundary bound
  state and additional charge excitations.}
\label{fig:plot9}
\end{figure}

In Fig.~\ref{fig:plot9} we plot $N^>_{\up}$ and $N^>_{\dw}$ as
functions of energy for $v_\mathrm{c}<v_\mathrm{s}$. We see that
the singularity arising due to the presence of a boundary bound state
appears only in the down-spin channel. For either spin polarization
we observe three dispersing features at $E_\mathrm{c}(q)$,
$E_\mathrm{s}(q)$ and $E_\mathrm{cs}(q)$ respectively. Their
interpretations are completely analogous to discussion in
Sec.~\ref{sec:LDOSattractive}.  In addition to these sharp peaks we
observe a broad maximum in the down-spin channel at energies $E\approx
E_\mathrm{bbs}+v_\mathrm{c}q/2$. This feature is suppressed for
$v_\mathrm{c}>v_\mathrm{s}$, see Fig.~\ref{fig:plot8}. Its physical
origin is the simultaneous excitation of the boundary bound state and
a finite energy excitation in the charge sector. 
We note that the asymmetry in  $N^>_\up(E,Q)-N^>_\dw(E,Q)$ could in
principle be detected in experiments using a magnetic STM tip.
So far we  have considered only hard-wall boundary conditions in the
charge sector, i.e. $\Phi_\mathrm{c}(x=0)=0$. Our analysis can be
straightforwardly extended to the case
$\Phi_\mathrm{c}(x=0)=\Phi_\mathrm{c}^0$, which
in terms of the original electrons corresponds to a local potential
close to the boundary.

\section{Finite-temperature LDOS}\label{sec:finitetemp}
Another interesting issue concerns the effects of a finite temperature on the
LDOS. The regime $T\alt\Delta/2$ can in principle be analyzed by generalizing
the methods recently developed in Refs.~[\onlinecite{finiteT}] to the boundary
state formalism. However, in order to keep matters simple we will restrict
ourselves to the regime of very low temperatures $T\ll\Delta$. Here the main
effects arise from a modification of the dynamical response in the gapless
charge sector and correlation functions in the spin sector can be approximated
by their $T=0$ expressions.  This is the case because we only consider
response functions that involve both sectors.  The charge part of the Green
function $G^{RL}_{\sigma\sigma'}(\tau,x_1,x_2)=
g_\mathrm{c}(\tau,x_1,x_2)\,g_\mathrm{s}(\tau,x_1,x_2)$ can be evaluated using
conformal field theory methods~\cite{Bloete-86,DiFrancescoMathieuSenechal97}
and is found to be
\begin{eqnarray}
g_\mathrm{c}(\tau,x_1,x_2)&=&-\frac{\delta_{\sigma\sigma'}}{2\pi}
\left(\frac{\pi}{v_\mathrm{c}\beta}\right)^{a+b}
\frac{1}{\sin^a\bigl(\tfrac{\pi}{v_\mathrm{c}\beta}
(v_\mathrm{c}\tau-2\ii R)\bigr)}\,
\frac{1}{\sin^b\bigl(\tfrac{\pi}{v_\mathrm{c}\beta}
(v_\mathrm{c}\tau+2\ii R)\bigr)}\nonumber\\*
& &\hspace{40mm}\times
\Biggl[\frac{\sinh\bigl(\tfrac{2\pi}{v_\mathrm{c}\beta}x_1\bigr)\,
\sinh\bigl(\tfrac{2\pi}{v_\mathrm{c}\beta}x_2\bigr)}
{\sin\bigl(\tfrac{\pi}{v_\mathrm{c}\beta}(v_\mathrm{c}\tau-\ii r)\bigr)\,
\sin\bigl(\tfrac{\pi}{v_\mathrm{c}\beta}(v_\mathrm{c}\tau+\ii r)\bigr)}
\Biggr]^c.
\label{eq:GRLchargefinteT}
\end{eqnarray}
Here $\beta=1/k_\mathrm{B}T$ and the exponents in the charge sector
were defined in \eqref{eq:exponents}. As we have already discussed,
the spin part $g_s(\tau,x_1,x_2$) is given by \eqref{eq:GRLspin}.

\begin{figure}[t]
\centering
\includegraphics[scale=0.34,clip=true]{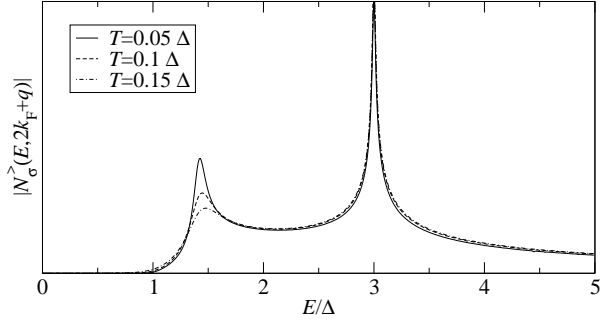}
\caption{$|N^>_\sigma(E,2k_\mathrm{F}+q)|$ (arbitrary units) for
  $v_\mathrm{s}q/\Delta=2$, $K_\mathrm{c}=K_\mathrm{s}=1$, and
  $v_\mathrm{c}=2v_\mathrm{s}$. We observe spectral weight within the spin gap
  and a broadening of the propagating peaks, which is much stronger for the
  spin peak at
  $E_\mathrm{s}=\sqrt{\left(\frac{v_\mathrm{s}q}{2}\right)^2+\Delta^2}$.}
\label{fig:plot10}
\end{figure}
\begin{figure}[t]
\centering
\includegraphics[scale=0.34,clip=true]{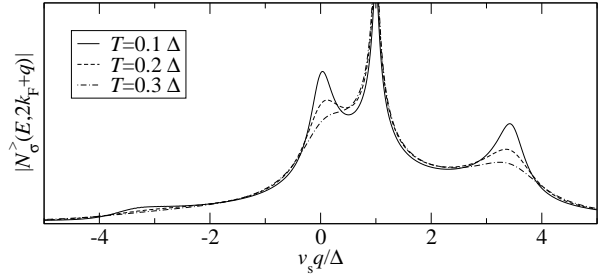}
\caption{Constant energy scan of $|N^>_\sigma(E,2k_\mathrm{F}+q)|$ (arbitrary 
  units) for $E=2\Delta$, $K_\mathrm{c}=K_\mathrm{s}=1$, and
  $v_\mathrm{c}=2v_\mathrm{s}$. We observe a suppression of the peak at $q=0$
  related to the pinned CDW.}
\label{fig:plot11}
\end{figure}
\begin{figure}[t]
\centering
\includegraphics[scale=0.34,clip=true]{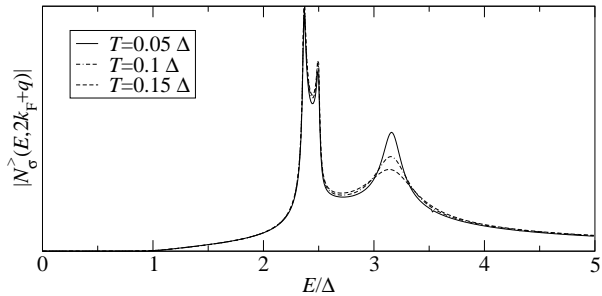}
\caption{$|N^>_\sigma(E,2k_\mathrm{F}+q)|$ (arbitrary units) for
  $v_\mathrm{s}q/\Delta=6$, $K_\mathrm{c}=K_\mathrm{s}=1$, and
  $v_\mathrm{s}=2v_\mathrm{c}$. We observe again that the broadening of the
  propagating spin peak is stronger than that of the other peaks.}
\label{fig:plot12}
\end{figure}
The particle contribution to the Fourier transform of the LDOS is still given
by \eqref{eq:defNp}. The form-factor expansion in the spin sector results in
the series expansion $N^>_\sigma(E,2k_\mathrm{F}+q)=\sum_{i}
N^>_i(E,2k_\mathrm{F}+q)$, where the first two terms can be cast in the form
\begin{eqnarray}
N^>_i(E,2k_\mathrm{F}+q)&=&
\frac{Z_1\,\pi^{a+b}}{32\pi^5v_\mathrm{c}}\,
\frac{e^{\ii\frac{\pi}{4}}\,e^{-\ii\frac{\pi}{2}(a+b+2c)}}
{(v_\mathrm{c}\beta)^{a+b-2}}\nonumber\\*
& &\times\int_{-\infty}^\infty d\theta\,h_i(\theta)
\int_{-\infty}^\infty dx \int_{-\infty}^x dy\,
\frac{e^{\frac{\ii\beta}{2\pi}(E_i+\frac{v_\mathrm{c}q_i}{2})x}}
{\sinh^a(x-\ii\delta)}\,
\frac{e^{\frac{\ii\beta}{2\pi}(E_i-\frac{v_\mathrm{c}q_i}{2})y}}
{\sinh^b(y-\ii\delta)}\,
\left(\frac{\sinh\bigl(\frac{1}{2}(y-x)\bigr)}
{\sinh\bigl(\frac{1}{2}(x+y-\ii\delta)\bigr)}\right)^{2c},
\label{eq:NiT}\\
& &\hspace{-20mm}h_1(\theta)=1,\quad 
h_2(\theta)=K^{\sigma\bar{\sigma}}\bigl(\theta+\ii\tfrac{\pi}{2}\bigr)\,
e^{\theta/2},\quad
E_1=E_2=E-\Delta\cosh\theta,\quad q_1=q,\quad
q_2=q-\frac{2\Delta}{v_\mathrm{s}}\sinh\theta.
\end{eqnarray}
We have plotted $N^>_\sigma(E,2k_\mathrm{F}+q)$ for different temperatures in
Figs.~\ref{fig:plot10}--\ref{fig:plot12}.  As expected a finite temperature
leads to a softening of the spectral gap $\Delta$, a suppression of the peak
related to the pinned CDW, and a broadening of the dispersing peaks. We
observe that the effect of an increasing temperature on the spin peak is much
stronger than the effect on the charge peak. The physical reason for this is
as follows: In the CDW state only the spin sector is protected by the gap.
Thus for $T\ll\Delta$ there exists a significant number of antiholons in the
thermal ground state.  They will participate in the distribution of the
external momentum $q$ after the creation of an additional antiholon-spinon
pair, thus leading to a decreased probability for the spinon to take the
momentum $q$ and thus to a suppression of the spin peak following
$E_\mathrm{s}(q)$. On the other hand, the charge peak is not affected as the
antiholons possess a linear dispersion.  This behavior is reminiscent of what
is found for the bulk spectral functions~\cite{Mottbulk}.

\section{Implications for STM experiments}\label{sec:exp}
STM experiments measure the local tunneling current, which is related to the
LDOS by Eq.~\eqref{eq:STM1}. In particular, the voltage dependence of the
tunneling conductance measures the thermally smeared LDOS of the sample at the
position of the tip. A possible spin dependence in the LDOS can be detected
using a magnetic tip.  As we have considered a one-dimensional model, our
results apply to quasi-1D materials at energies above the 1D-3D cross-over
scale, which is set by the strength of the 3D couplings. Furthermore, the main
feature of the model we have studied is the existence of a spectral gap in one
of the sectors, while the excitations in the other sector remain gapless. This
situation is experimentally realized in various materials, for example in
two-leg ladder materials~\cite{Tennant}, stripe phases of
HTSC~\cite{Fischer-07,Kivelson-03}, carbon
nanotubes~\cite{Odom-02,Wildoer-98}, Bechgaard salts~\cite{Bechgaard}, and
chain materials~\cite{1DMott} like ${\rm SrCuO_2}$ and ${\rm Sr_2CuO_3}$. As
our results show, STM experiments can be used to extract rather detailed
information regarding bulk excitations by analyzing the modification of the
LDOS due to a boundary/impurity.

Perhaps the most interesting materials to which our findings may be applied
on a qualitative level are two-leg ladders like~\cite{Abbamonte-04}
Sr$_{14}$Cu$_{24}$O$_{41}$ which possesses a spin
gap~\cite{Magishi-98} of $\Delta\approx 550K$.  The model we have
studied captures the most basic features of the low-energy description
of (weakly doped) two-leg ladders, namely a gapless charge sector and
a gapped spin sector. While the description of the spin sector for
weakly doped two-leg ladders is considerably more involved, we expect
the gross features to be similar. In particular, we expect peaks to
appear in $N_\sigma(E,Q)$, which correspond to the pinned CDW order,
dispersing spin and charge degrees of freedom, and possibly boundary
bound states. We note that our results apply to the regime $T\ll
E,\Delta,v_\mathrm{c}/a_0$, where temperature effects are negligible. 

In Ref.~[\onlinecite{Kivelson-03}] it was proposed that STM and STS
experiments in HTSC can be used to detect ``fluctuating stripes'' (i.e.
incommensurate spin and charge fluctuations on energy scales small compared to
the superconducting gap) by rendering them static by the effects of impurities
with a potential comparable to the (low) energy scales of these fluctuations.
In that work it was also argued that 1D Luttinger liquids are effectively
quantum critical systems and that a form of local (power law) CDW order is
effectively induced by impurities (and edges) which pin the phase of the CDW,
an effect that we have shown here to take place in a 1D Luther-Emery liquid
associated with a CDW. STM and STS experiments in {\BSCCO} have confirmed the
existence of both non-dispersive spectral features in the LDOS associated with
``fluctuating stripe order'' as well as dispersive features associated with
the propagating quasiparticles of the
superconductor\cite{Howald-01,Kivelson-03,Kohsaka-07}. Recent STS experiments
in {\BSCCO} have shown that the dispersive features of the LDOS disappear
above the $T_\mathrm{c}$ of the superconductor while the non-dispersive
features survive up to the temperature $T^*$ at which the pseudogap
closes\cite{Yazdani-10}.

\section{Conclusions}
In this work we have determined the spatial Fourier transform of the
LDOS of one-dimensional CDW states and Mott insulators in presence
of a boundary. The latter may either model a strong potential impurity
or be realized in a two-tip STM experiment. 
We found that the Fourier transform of the LDOS is dominated by a
singularity at an energy equal to the single-particle gap $\Delta$ and 
at momentum $2k_\mathrm{F}$. This feature is indicative of the pinning
of the CDW order at the position of the impurity. We observed clear
signatures of dispersing spin and charge excitations, which can be
used to infer the nature of the underlying electron-electron
interactions. In the case of CDW states with repulsive interactions we
find a spin mode and a linear dispersing charge mode while for
attractive interactions a third dispersing mode appears, which can be
thought of as arising from a spin excitation with a fixed momentum
$q_0$ and a charge excitation with momentum $q-q_0$.

We have also investigated the modification of the LDOS due to boundary
bound states. These may arise in presence of boundary potentials or
magnetic fields. We found that boundary bound states give rise to
non-dispersing singularities at energies below the single-particle
gap. While the bound state is formed in the gapped sector of the
theory, the exponent of the corresponding singularity only depends
on the Luttinger liquid parameter of the gapless sector. 
We have analyzed temperature effects in regime $T\ll \Delta$ and
discussed implications of our results for STM measurements on
quasi-1D materials such as doped two-leg ladders.

\section*{Acknowledgments}
We would like to thank Joe Bhaseen, Dmitry Kovrizhin, and Christian Pfleiderer
for useful discussions.  DS was supported by the Deutsche Akademie der
Naturforscher Leopoldina under grant BMBF-LPD 9901/8-145 while working at the
Rudolf Peierls Centre for Theoretical Physics, University of Oxford.  This
work was also supported by the EPSRC under grant EP/D050952/1 (FHLE), the NSF
under grant DMR 0758462 at the University of Illinois (EF), the U.S.
Department of Energy, Division of Materials Sciences under Award No.
DE-FG02-07ER46453 through the Frederick Seitz Materials Research Laboratory of
the University of Illinois (EF, AJ) and the ESF network INSTANS.

\appendix
\section{Renormalization group analysis of an impurity potential}\label{sec:RG}
Let us consider the low-energy theory of a one-dimensional CDW state on the
infinite line $H=H_\mathrm{c}+H_\mathrm{s}$, where 
\begin{eqnarray}
H_\mathrm{c}&=&\frac{v_\mathrm{c}}{16\pi}\int_{-\infty}^\infty dx 
\biggl[\frac{1}{K_\mathrm{c}^2}\bigr(\partial_x\Phi_\mathrm{c}\bigr)^2+
K_\mathrm{c}^2\bigr(\partial_x\Theta_\mathrm{c}\bigr)^2\biggr],\\
H_\mathrm{s}&=&\frac{v_\mathrm{s}}{16\pi}\int_{-\infty}^\infty dx 
\biggl[\frac{1}{K_\mathrm{s}^2}\bigr(\partial_x\Phi_\mathrm{s}\bigr)^2+
K_\mathrm{s}^2\bigr(\partial_x\Theta_\mathrm{s}\bigr)^2\biggr]
-\frac{g_\mathrm{s}}{(2\pi)^2}\int_{-\infty}^\infty dx\, \cos\Phi_\mathrm{s}.
\label{eq:RGspinham}
\end{eqnarray}
We want to study the effect of an impurity potential at position $x=0$, which
in bosonized form reads
\begin{equation}
V_\mathrm{imp}=\lambda\int_{-\infty}^\infty dx\,\delta(x)\,
\cos\left(\frac{\Phi_\mathrm{c}}{2}\right)
\cos\left(\frac{\Phi_\mathrm{s}}{2}\right).
\label{eq:RGimppot}
\end{equation}
As the spin sector in the bulk is massive, we have
$\big\langle\!\cos(\Phi_\mathrm{s}/2)\big\rangle\neq 0$, which implies that at
low energies we can approximate $\cos\bigl(\Phi_\mathrm{c}/2\bigr)
\cos\bigl(\Phi_\mathrm{s}/2\bigr)$ in \eqref{eq:RGimppot} by
$\big\langle\!\cos\bigl(\Phi_\mathrm{c}/2\bigr)\big\rangle
\cos\bigl(\Phi_\mathrm{s}/2\bigr)+
\cos\bigl(\Phi_\mathrm{c}/2\bigr)
\big\langle\!\cos\bigl(\Phi_\mathrm{s}/2\bigr)\big\rangle$.
Thus in the charge sector we get a boundary sine-Gordon model~\cite{Gogolin}.
For $K_\mathrm{c}^2<2$ the impurity scattering potential scales to strong
coupling.  Hence as long as the interactions are not too attractive the field
$\Phi_\mathrm{c}$ gets pinned at the boundary, $\Phi_\mathrm{c}(0)=0$. This in
turn induces an impurity contribution in the gapped spin sector
\begin{equation}
V_\mathrm{imp,s}=\lambda\,
\left\langle\cos\left(\frac{\Phi_\mathrm{c}(0)}{2}\right)\right\rangle\;
\int_{-\infty}^\infty dx\,\delta(x)\,
\cos\left(\frac{\Phi_\mathrm{s}}{2}\right).
\label{eq:RGimppots}
\end{equation}
If one analyzes the bulk and boundary cosine terms in the resulting impurity
model \eqref{eq:RGspinham} and \eqref{eq:RGimppots} simultaneously, the
leading order renormalization group equations are given by the scaling
dimensions of the perturbing operators
\begin{equation}
\frac{d g_\mathrm{s}}{dl}=2\,\bigl(1-K_\mathrm{s}^2\bigr)\,g_\mathrm{s},\quad
\frac{d\lambda}{dl}=\left(1-\frac{K_\mathrm{s}^2}{2}\right)\lambda.
\end{equation}
As long as $K_\mathrm{s}^2>2/3$ the boundary term grows more rapidly than the
bulk term. Assuming that it reaches the strong-coupling regime first leads to
the pinning of the spin field $\Phi_\mathrm{s}(0)=0$. This cuts the chain in
two half-lines and we obtain the model \fr{eq:hamiltonian}--\fr{hardwall}.

\section{Calculation of the Green function: Charge sector}
\label{sec:appgfcs}
The Green function \eqref{eq:GF} factorizes into a product of correlation
functions in the spin and charge sectors. For example, using the bosonization
identities \eqref{eq:bosonizationR} and \eqref{eq:bosonizationL} the
$2k_\mathrm{F}$-component $G^{RL}_{\sigma\sigma'}$ can be written as
\begin{equation}
G^{RL}_{\sigma\sigma'}(\tau,x_1,x_2)=
-\frac{1}{2\pi}\,
\Big\langle e^{-\frac{\ii}{2}\phi_\mathrm{c}(\tau,x_1)}\,
e^{-\frac{\ii}{2}\bar{\phi}_\mathrm{c}(0,x_2)}\Big\rangle_\mathrm{c}\,
\Big\langle e^{-\frac{\ii}{2}f_\sigma\phi_\mathrm{s}(\tau,x_1)}\,
e^{-\frac{\ii}{2}f_{\sigma'}\bar{\phi}_\mathrm{s}(0,x_2)}
\Big\rangle_\mathrm{s}.
\label{eq:GFboson}
\end{equation}
Both correlation functions have to be determined in the presence of the
boundary at $x=0$. The charge part is calculated below using a
standard mode expansion~\cite{FabrizioGogolin95}, the spin part will
be calculated in App.~\ref{sec:appgfss}. 

In order to obtain the correlation functions in the charge sector we first
bring the Hamiltonian \eqref{eq:chargehamiltonian} to standard form by
rescaling the fields as $\Phi_\mathrm{c}\rightarrow
K_\mathrm{c}\Phi_\mathrm{c}$,
$\Theta_\mathrm{c}\rightarrow\Theta_\mathrm{c}/K_\mathrm{c}$. The charge parts
of the operators \eqref{eq:bosonizationR} and \eqref{eq:bosonizationL} then
become
\begin{eqnarray}
\exp\Bigl(\pm\tfrac{\ii}{2}\phi_\mathrm{c}(\tau,x)\Bigr)
&\rightarrow&e^{\ii\pi sc/4}
\exp\Bigl(\pm\tfrac{\ii}{2}c\,\phi_\mathrm{c}(z)\Bigr)
\exp\Bigl(\pm\tfrac{\ii}{2}s\,\bar{\phi}_\mathrm{c}(\bar{z})\Bigr)=
e^{-\ii\pi sc/4}
\exp\Bigl(\pm\tfrac{\ii}{2}s\,\bar{\phi}_\mathrm{c}(\bar{z})\Bigr)
\exp\Bigl(\pm\tfrac{\ii}{2}c\,\phi_\mathrm{c}(z)\Bigr),
\label{eq:rescaledL}\\*
\exp\Bigl(\pm\tfrac{\ii}{2}\bar{\phi}_\mathrm{c}(\tau,x)\Bigr)
&\rightarrow&e^{\ii\pi sc/4}
\exp\Bigl(\pm\tfrac{\ii}{2}s\,\phi_\mathrm{c}(z)\Bigr)
\exp\Bigl(\pm\tfrac{\ii}{2}c\,\bar{\phi}_\mathrm{c}(\bar{z})\Bigr)=
e^{-\ii\pi sc/4}
\exp\Bigl(\pm\tfrac{\ii}{2}c\,\bar{\phi}_\mathrm{c}(\bar{z})\Bigr)
\exp\Bigl(\pm\tfrac{\ii}{2}s\,\phi_\mathrm{c}(z)\Bigr),
\label{eq:rescaledR}
\end{eqnarray}
where we have already used \eqref{eq:commphibarphi} and assumed $-L<x<0$. The
constants are parameterized via $s=\sinh\xi_\mathrm{c}$ and
$c=\cosh\xi_\mathrm{c}$ with $K_\mathrm{c}=e^{\xi_\mathrm{c}}$. The complex
coordinates are defined as $z=v_\mathrm{c}\tau-\ii x$,
$\bar{z}=v_\mathrm{c}\tau+\ii x$. The charge part of the Green function can
hence be obtained from the four-point function
\begin{equation}
\Big\langle 
e^{\ii \beta_1\bar{\phi}_\mathrm{c}(\bar{z}_1)}\,
e^{\ii \alpha_1\phi_\mathrm{c}(z_1)}\,
e^{\ii \alpha_2\phi_\mathrm{c}(z_2)}\,
e^{\ii \beta_2\bar{\phi}_\mathrm{c}(\bar{z}_2)}
\Big\rangle_\mathrm{UHP},
\label{eq:cfcharge}
\end{equation}
where $\alpha_{1,2},\beta_{1,2}\in\mathbb{R}$, $z_1=v_\mathrm{c}\tau-\ii x_1$
and $z_2=-\ii x_2$ lie in the upper half-plane.

We calculate \eqref{eq:cfcharge} from the mode expansions for the chiral
fields $\phi_\mathrm{c}$ and $\bar{\phi}_\mathrm{c}$. These are obtained by
first noting that the fields $\Phi_\mathrm{c}$ and $\Theta_\mathrm{c}$ have to
satisfy the equations of motion
$v_\mathrm{c}\partial_x\Theta_\mathrm{c}=-\ii\partial_\tau\Phi_\mathrm{c}$ and
$\partial_\tau\Theta_\mathrm{c}=\ii v_\mathrm{c}\partial_x\Phi_\mathrm{c}$ as
well as the boundary conditions
$\Phi_\mathrm{c}(x=0)=\Phi_\mathrm{c}(x=-L)=0$. The semi-infinite system is
obtained by taking $L\rightarrow\infty$. This yields the mode expansions
\begin{eqnarray}
\Phi_\mathrm{c}(\tau,x)&=&-
\frac{x}{L}\hat{\Pi}_0
+\ii\sum_{n=1}^\infty\frac{\sin\frac{n\pi x}{L}}{\sqrt{n\pi}}
\left(b_n\,e^{-n\pi v_\mathrm{c}\tau/L}-
b_n^\dagger\,e^{n\pi v_\mathrm{c}\tau/L}\right),\label{eq:modephi}\\*
\Theta_\mathrm{c}(\tau,x)&=&\hat{\Theta}_0-
\ii\,\frac{v_\mathrm{c}\tau}{L}\hat{\Pi}_0
+\sum_{n=1}^\infty\frac{\cos\frac{n\pi x}{L}}{\sqrt{n\pi}}
\left(b_n\,e^{-n\pi v_\mathrm{c}\tau/L}+
b_n^\dagger\,e^{n\pi v_\mathrm{c}\tau/L}\right),\label{eq:modetheta}
\end{eqnarray}
where the zero-mode operator $\hat{\Pi}_0$ has the discrete spectrum $2\pi m$,
$m\in\mathbb{Z}$, and $\comm{\hat{\Theta}_0}{\hat{\Pi}_0}=8\pi\ii$,
$\comm{b_m}{b_n^\dagger}=8\pi\,\delta_{mn}$.  The mode expansions for the
chiral fields are easily obtained via
$\phi_\mathrm{c}=(\Phi_\mathrm{c}+\Theta_\mathrm{c})/2$ and
$\bar{\phi}_\mathrm{c}=(\Phi_\mathrm{c}-\Theta_\mathrm{c})/2$. Their
commutation relations are
\begin{equation}
\bigl[\phi_\mathrm{c}(\tau,x),\bar{\phi}_\mathrm{c}(\tau,x')\bigr]=
\left\{\begin{array}{lll}
0&,&x=x'=0,\\
4\pi\ii&,&x=x'=-L,\\
2\pi\ii&,&\text{else},
\end{array}\right.\label{eq:commphibarphi}
\end{equation}
as well as
$\bigl[\phi_\mathrm{c}(\tau,x),\phi_\mathrm{c}(\tau,x')\bigr]=
-\bigl[\bar{\phi}_\mathrm{c}(\tau,x),\bar{\phi}_\mathrm{c}(\tau,x')\bigr]=
2\pi\ii\,\sgn{x-x'}$, where $\sgn{0}=0$. Similar mode expansions were
obtained in Refs.~[\onlinecite{FabrizioGogolin95}]. Given the mode
  expansion it is straightforward to calculate the four-point function
  \fr{eq:cfcharge}. We find
\begin{equation}
\Big\langle 
e^{\ii \beta_1\bar{\phi}_\mathrm{c}(\bar{z}_1)}\,
e^{\ii \alpha_1\phi_\mathrm{c}(z_1)}\,
e^{\ii \alpha_2\phi_\mathrm{c}(z_2)}\,
e^{\ii \beta_2\bar{\phi}_\mathrm{c}(\bar{z}_2)}
\Big\rangle_\mathrm{UHP}=
\frac{C\delta_{\alpha_1+\alpha_2,\beta_1+\beta_2}\
(z_1-z_2)^{2\alpha_1\alpha_2}\,(\bar{z}_1-\bar{z}_2)^{2\beta_1\beta_2}}
{(\bar{z}_1-z_1)^{2\alpha_1\beta_1}(\bar{z}_1-z_2)^{2\alpha_2\beta_1}
(z_1-\bar{z}_2)^{2\alpha_1\beta_2}(z_2-\bar{z}_2)^{2\alpha_2\beta_2}}\ ,
\label{eq:correlationschargesector}
\end{equation}
where $C\in\mathbb{R}$ is a constant which we set to one throughout this
manuscript.  This result implies Eq.~\eqref{eq:GRLcharge}. The
finite-temperature correlation functions are
obtained~\cite{Bloete-86,DiFrancescoMathieuSenechal97} by mapping
\eqref{eq:correlationschargesector} onto a cylinder of circumference
$v_\mathrm{c}/k_\mathrm{B}T$.

\section{Calculation of the Green function: Spin sector}
\label{sec:appgfss}
The calculation of the correlation functions in the spin sector relies on the
integrability of the sine-Gordon model on the half-line.  We use the boundary
state formalism introduced by Ghoshal and
Zamolodchikov~\cite{GhoshalZamolodchikov94} together with a form-factor
expansion based on form factors obtained by Lukyanov and
Zamolodchikov~\cite{LukyanovZamolodchikov01}. The analogous expansion for the
quantum Ising model has been analyzed in Ref.~[\onlinecite{SE07}].  We will
first discuss the general formalism and then derive \eqref{eq:GRLspin}.

\subsection{Boundary state formalism and form-factor expansion}
\begin{figure}[t]
\psfrag{REP1}{$Z_{a_1}^\dagger(\theta_1)$}
\psfrag{REP1a}{$Z_{b_1}^\dagger(\theta_1)$}
\psfrag{REP2}{$Z_{a_2}^\dagger(\theta_2)$}
\psfrag{REP2a}{$Z_{b_2}^\dagger(\theta_2)$}
\psfrag{REP3}{$S_{a_1a_2}^{b_1b_2}(\theta_{1}-\theta_{2})$}
\psfrag{REP4}{$Z_a^\dagger(\theta)$}
\psfrag{REP4a}{$Z_b^\dagger(-\theta)$}
\psfrag{REP5}{$R_a^b(\theta)$}
\includegraphics[scale=0.2]{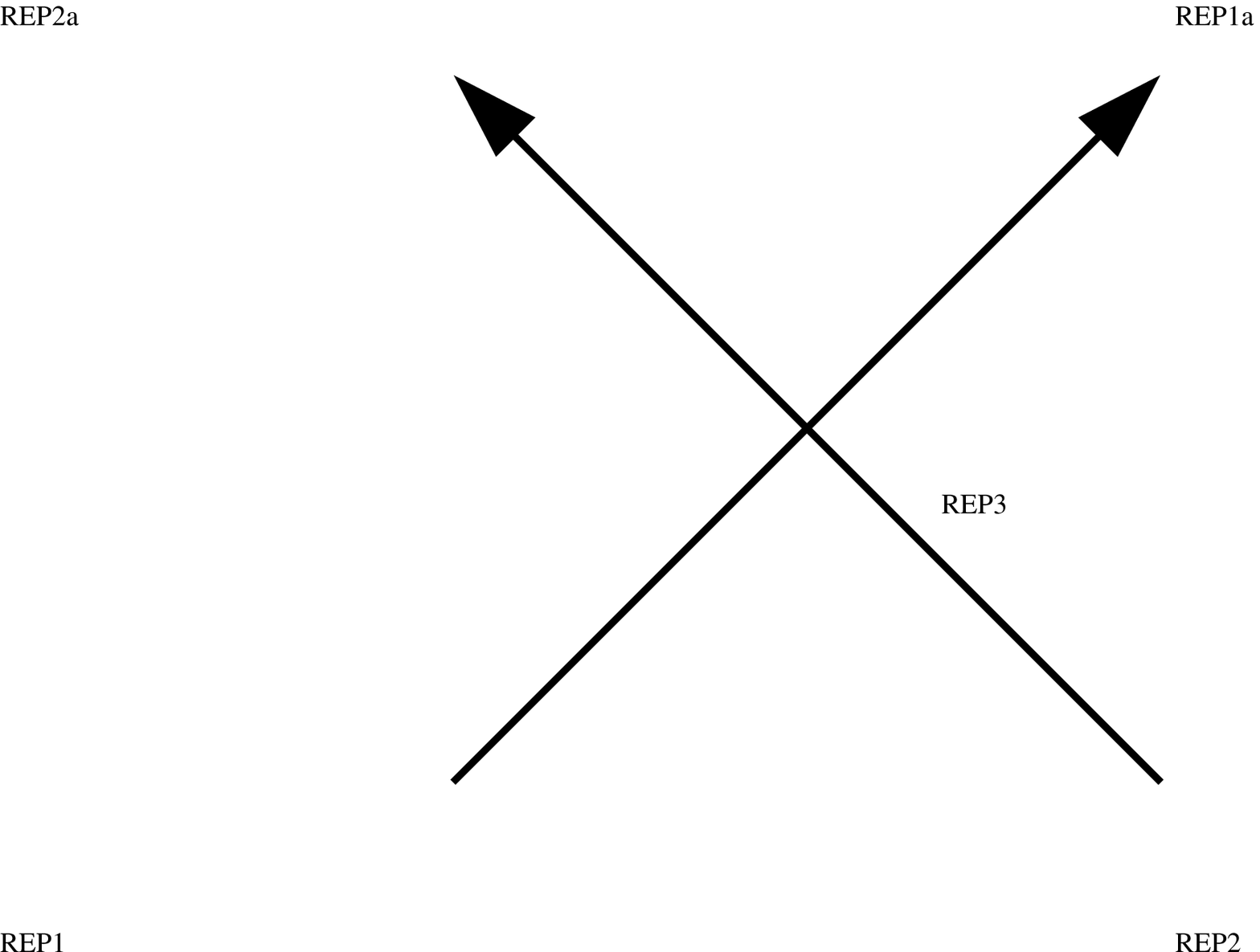}\hspace{30mm}
\includegraphics[scale=0.2]{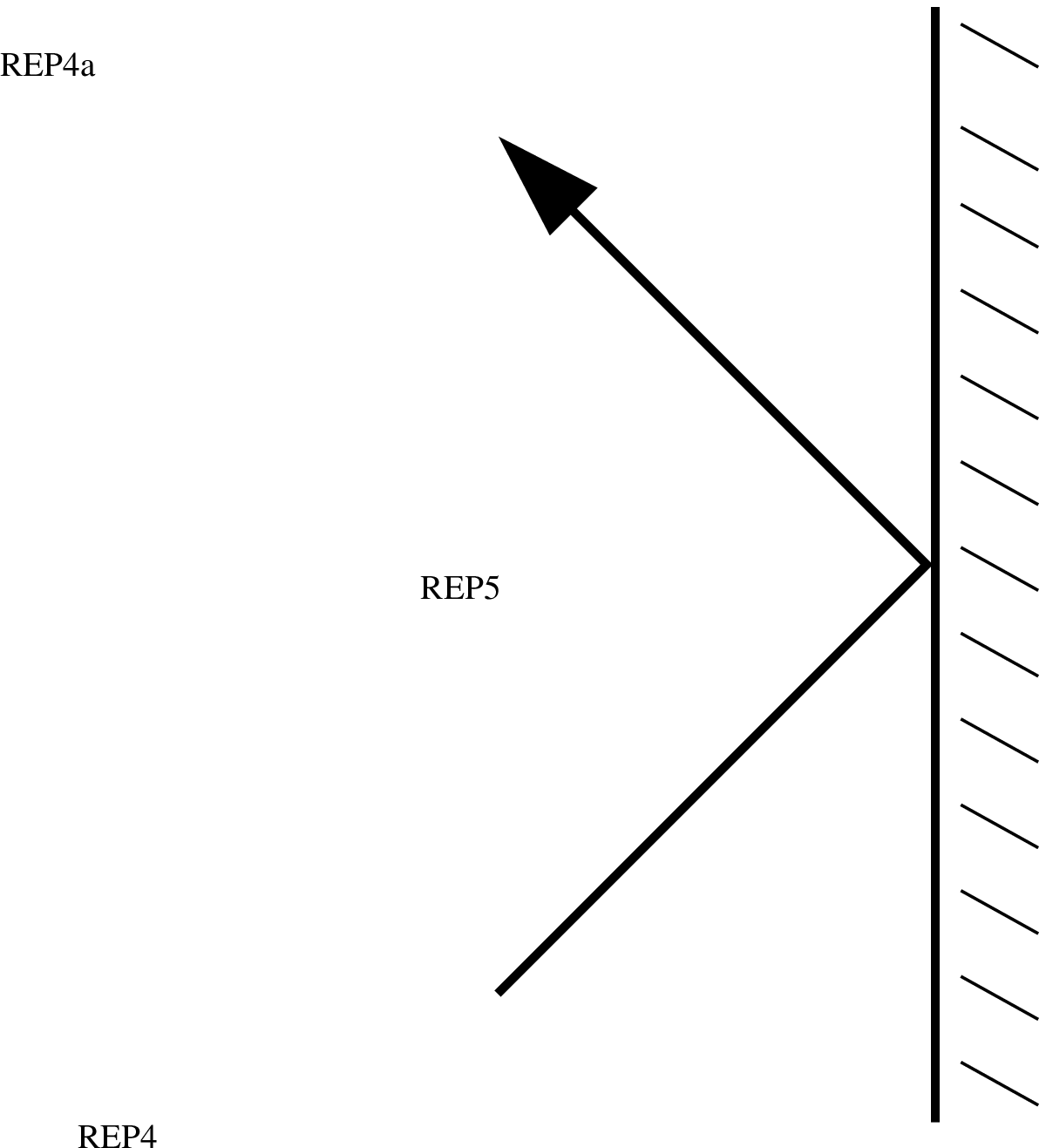}
\caption{Two-particle scattering and scattering off the boundary.}
\label{fig:scattering}
\end{figure}
Let us consider the sine-Gordon model \eqref{eq:spinhamiltonian} in
the half-plane $(\tau,x)$, $\tau\in\mathbb{R}$, $-\infty<x\le 0$. The
boundary is located at $x=0$ and $\tau$ denotes imaginary time
($\tau=\ii t$).  The Hilbert space of states associated with the
semi-infinite line $\tau=\text{const}$, $-\infty<x\le 0$, is denoted
by $\mathcal{H}_\mathrm{b}$. We obtain the Euclidean action in its
standard form by rescaling the fields according to
$\Phi_\mathrm{s}\rightarrow
\Phi_\mathrm{s}'=\Phi_\mathrm{s}/K_\mathrm{s}$ and
$\Theta_\mathrm{s}\rightarrow
\Theta_\mathrm{s}'=K_\mathrm{s}\Theta_\mathrm{s}$.  The action of the
sine-Gordon model with a boundary is then given
by~\cite{GhoshalZamolodchikov94} (we set $v_\mathrm{s}=1$)
\begin{equation}
\mathcal{S}_\mathrm{bsG}=\frac{1}{16\pi}\int d\tau \int_{-\infty}^0 dx 
\biggl[\bigl(\partial_\tau\Phi_\mathrm{s}'\bigr)^2+
\bigl(\partial_x\Phi_\mathrm{s}'\bigr)^2
-\frac{4 g_\mathrm{s}}{\pi}\cos\bigl(K_\mathrm{s}\Phi_\mathrm{s}'\bigr)
\biggr]
-g_\mathrm{b}\int d\tau\,
\cos\left(\frac{K_\mathrm{s}}{2}
\bigl(\Phi_\mathrm{s}'\big|_{x=0}-\Phi_\mathrm{s}^0/K_\mathrm{s}\bigr)\right),
\label{eq:IFTboundary}
\end{equation}
where $g_\mathrm{s}$, $g_\mathrm{b}$ and $\Phi_\mathrm{s}^0$ are free
parameters.  (We use the conventions $0<K_\mathrm{s}<1$, the action as given
in Ref.~[\onlinecite{GhoshalZamolodchikov94}] is obtained by another rescaling
of the fields by $\sqrt{8\pi}$.) The cases $g_\mathrm{b}=0$ and
$g_\mathrm{b}\rightarrow\infty$ correspond to free and fixed boundary
conditions, respectively.  We stress that in the case of fixed boundary
conditions $\Phi_\mathrm{s}'(x=0)=\Phi_\mathrm{s}^0/K_\mathrm{s}$ implies
$\Phi_\mathrm{s}(x=0)=\Phi_\mathrm{s}^0$ in the original system
\eqref{eq:spinhamiltonian}.  As was conjectured by Ghoshal and
Zamolodchikov~\cite{GhoshalZamolodchikov94} and shown
independently~\cite{MacIntyre95} by MacIntyre and Saleur, Skorik and Warner,
the classical sine-Gordon model on the half-line \eqref{eq:IFTboundary}
possesses infinitely many integrals of motion and is hence integrable.

We start by summarizing some results for the \emph{bulk} sine-Gordon
model, i.e. the theory without boundary. In the repulsive regime
($K_\mathrm{s}>1/\sqrt{2}$) a basis of the Hilbert space $\mathcal{H}$
is given by scattering states of solitons and antisolitons
\begin{equation}
  \ket{\theta_1,\ldots,\theta_n}_{a_1,\ldots,a_n}=
  Z^\dagger_{a_1}(\theta_1)\ldots Z^\dagger_{a_n}(\theta_n)\ket{0},\quad
  ^{a_n,\ldots,a_1}\!\!\bra{\theta_n,\ldots,\theta_1}=
  \bra{0}Z_{a_n}(\theta_n)\ldots Z_{a_1}(\theta_1),
  \label{eq:defA}
\end{equation}
where $a_i=\pm 1$ and $\ket{0}$ is the ground state in absence of a boundary.
Solitons and antisolitons are created by the operators $Z^\dagger_-(\theta)$
and $Z^\dagger_+(\theta)$. They are characterized by a topological U(1) charge
$-1$ and $1$, respectively, while their energy and momentum are parametrized
in terms of the rapidity $\theta$ by $E=\Delta\cosh\theta$ and
$P=\Delta\sinh\theta$.  The dependence of the soliton mass $\Delta$ on the
bare parameters in the action was obtained in
Ref.~[\onlinecite{Zamolodchikov95}]. We note that in the attractive regime
($K_\mathrm{s}<1/\sqrt{2}$) breather (soliton-antisoliton) bound states occur
as well.  The operators $Z_a$ and $Z_a^\dagger$ fulfill the
Faddeev--Zamolodchikov algebra~\cite{ZamolodchikovZamolodchikov79} (see
Fig.~\ref{fig:scattering})
\begin{eqnarray}
Z_{a_1}(\theta_1)Z_{a_2}(\theta_2)&=&S_{a_1a_2}^{b_1b_2}(\theta_1-\theta_2)
Z_{b_2}(\theta_2)Z_{b_1}(\theta_1),\nonumber\\*
Z_{a_1}^\dagger(\theta_1)Z_{a_2}^\dagger(\theta_2)&=&
S_{a_1a_2}^{b_1b_2}(\theta_1-\theta_2)
Z_{b_2}^\dagger(\theta_2)Z_{b_1}^\dagger(\theta_1),\label{eq:Aalgebra}\\*
Z_{a_1}(\theta_1)Z_{a_2}^\dagger(\theta_2)&=&
2\pi\delta(\theta_1-\theta_2)\delta_{a_1a_2}
+S_{a_2b_1}^{b_2a_1}(\theta_1-\theta_2)
Z_{b_2}^\dagger(\theta_2)Z_{b_1}(\theta_1).\nonumber
\end{eqnarray}
Here $S_{a_1a_2}^{b_1b_2}(\theta)$ is the two-particle scattering matrix,
which was derived in
Refs.~[\onlinecite{Zamolodchikov77,korepinSmat}]. The unitarity
condition reads
$S_{a_1a_2}^{c_1c_2}(\theta)S_{c_1c_2}^{b_1b_2}(-\theta)=
\delta_{a_1}^{b_1}\delta_{a_2}^{b_2}$.  Its
non-vanishing elements are
\begin{equation}
\label{eq:Smatrix}
S^{++}_{++}(\theta)=S^{--}_{--}(\theta),\quad
S^{+-}_{+-}(\theta)=S^{-+}_{-+}(\theta),\quad
S^{+-}_{-+}(\theta)=S^{-+}_{+-}(\theta),
\end{equation}
for which explicit expressions can be found for example in
Ref.~[\onlinecite{EsslerKonik05}].  At the LEP ($K_\mathrm{s}=1/\sqrt{2}$) the
scattering matrix simplifies to
$S_{a_1a_2}^{b_1b_2}(\theta)=-\delta_{a_1}^{b_1}\delta_{a_2}^{b_2}$, while in 
the spin symmetric case ($K_\mathrm{s}=1$) one has~\cite{MattssonDorey00}
\begin{equation}
  S^{++}_{++}(\theta)=S_0(\theta)\equiv
  -\frac{\Gamma\bigl(1+\tfrac{\ii\theta}{2\pi}\bigr)\,
    \Gamma\bigl(\tfrac{1}{2}-\tfrac{\ii\theta}{2\pi}\bigr)}
  {\Gamma\bigl(1-\tfrac{\ii\theta}{2\pi}\bigr)\,
    \Gamma\bigl(\tfrac{1}{2}+\tfrac{\ii\theta}{2\pi}\bigr)},\quad
  S^{+-}_{+-}(\theta)=-\frac{\theta}{\theta-\ii\pi}\,S_0(\theta),\quad
  S^{+-}_{-+}(\theta)=-\frac{\ii\pi}{\theta-\ii\pi}\,S_0(\theta).
\end{equation}
We note that the Faddeev--Zamolodchikov algebra \eqref{eq:Aalgebra} is
invariant under the unitary transformation $Z_a(\theta)\rightarrow
e^{\ii\varphi}Z_a(\theta)$, which changes the basis of scattering states. In
terms of the basis states \eqref{eq:defA} the resolution of the identity reads
\begin{equation}
\mathrm{id}=\ket{0}\bra{0}+\sum_{n=1}^\infty\frac{1}{n!}\sum_{\{a_i\}}
\int_{-\infty}^\infty\frac{d\theta_1\ldots d\theta_{n}}{(2\pi)^n}
\ket{\theta_n,\ldots,\theta_1}_{a_n,\ldots,a_1}\,\!\!
^{a_1,\ldots,a_n}\!\!\bra{\theta_1,\ldots,\theta_n}.
\label{eq:residentity}
\end{equation}

The boundary can be introduced\cite{GhoshalZamolodchikov94} as an
infinitely heavy, impenetrable particle $B$ sitting at $x=0$. The
ground state in presence of the boundary can then be represented as
$\ket{0_\mathrm{b}}=B\ket{0}$. Scattering of elementary excitations
off the boundary is encoded in the relations (see
Fig.~\ref{fig:scattering})
\begin{equation}
\label{eq:boundaryscattering}
Z_a^\dagger(\theta)B=R_a^b(\theta)Z_b^\dagger(-\theta)B,
\end{equation}
where the functions $R_a^b(\theta)$ are the single-particle reflection
amplitudes. In order to preserve integrability, the boundary
reflection matrix $R(\theta)$ has to satisfy a number of conditions
which were discussed in Ref.~[\onlinecite{GhoshalZamolodchikov94}]. At
the LEP and for Dirichlet boundary conditions
$\Phi_\mathrm{s}(\tau,x=0)=\Phi_\mathrm{s}^0$,
$0\le\Phi_\mathrm{s}^0<\pi$ in the original system
\eqref{eq:spinhamiltonian}, it is given by~\cite{Ameduri-95} 
\begin{equation}
R_\pm^\pm(\theta)=-
\frac{\cosh\Bigl(\ii\frac{\pi}{4}\pm\ii\frac{\Phi_\mathrm{s}^0}{2}
+\frac{\theta}{2}\Bigr)}
{\cosh\left(\ii\frac{\pi}{4}\pm\ii\frac{\Phi_\mathrm{s}^0}{2}
-\frac{\theta}{2}\right)},
\quad R_\pm^\mp(\theta)=0.
\label{eq:RLE}
\end{equation}
For $\pi/2\le\Phi_\mathrm{s}^0$ the reflection amplitude $R^+_+$ possesses a
simple pole in the physical strip $0\le\mathfrak{Im}\,\theta\le\pi/2$, which
indicates the existence of a boundary bound state.  The overall sign of the
reflection matrix is fixed by the requirement
$-\ii\,\text{Res}\bigl[R^\pm_\pm(\theta),
\theta=\pm\ii\,(\Phi_\mathrm{s}^0\mp\pi/2)\bigr]=-2\cos\Phi_\mathrm{s}^0>0$,
see App.~\ref{sec:appbbs} below.  Explicit representations of $R$ for general
$K_\mathrm{s}$ can be found in
Refs.~[\onlinecite{GhoshalZamolodchikov94,MattssonDorey00,Caux-03}].  For
$K_\mathrm{s}=1$ and $\Phi_\mathrm{s}^0=0$ one finds in particular
\begin{equation}
\label{eq:RSU2inv}
R_\pm^\pm(\theta)=
-\frac{\Gamma\bigl(1+\tfrac{\ii\theta}{2\pi}\bigr)\,
\Gamma\bigl(\tfrac{1}{2}-\tfrac{\ii\theta}{\pi}\bigr)}
{\sqrt{\pi}\,\Gamma\bigl(1-\tfrac{\ii\theta}{2\pi}\bigr)}\,
2^{\frac{\ii}{\pi}\theta}\,
\left(\cosh\frac{\theta}{2}+\ii\sinh\frac{\theta}{2}\right),
\quad R_\pm^\mp(\theta)=0.
\end{equation}
The vanishing of the off-diagonal amplitudes $R_\pm^\mp(\theta)=0$ is
a consequence of fixed boundary conditions and holds for general
$K_\mathrm{s}$.

Our aim is to calculate the time-ordered two-point function
\begin{equation}
\label{eq:cfboundary}
C(\tau,x_1,x_2)=
\bra{0_\mathrm{b}}\mathcal{T}_\tau\,O_1(\tau,x_1)\,
O_2(0,x_2)\ket{0_\mathrm{b}}.
\end{equation}
Here the time-dependence of the operators is given by $O_i(\tau,x)=
e^{\tau H_\mathrm{b}}\,O_i(0,x)\,e^{-\tau H_\mathrm{b}}$, where
$H_\mathrm{b}$ is the Hamiltonian of the system in the presence of the
boundary \eqref{eq:spinhamiltonian}. Given that in the Euclidean
formalism $\tau$ and $x$ are interchangeable one may equally well
designate $x$ to be the ``Euclidean time''. In this picture the
equal-time section is the infinite line, $x=\text{const}$,
$-\infty<\tau<\infty$, and the associated Hilbert space $\mathcal{H}$
is that of the corresponding \emph{bulk theory}.  The boundary at
$x=0$ now appears as an initial condition which is expressed in terms
of a ``boundary state'' $\ket{\mathrm{B}}$. It was shown by Ghoshal
and Zamolodchikov~\cite{GhoshalZamolodchikov94} that the correlation
function \eqref{eq:cfboundary} can be expressed as
\begin{equation}
\label{eq:cfboundarystate}
C(\tau,x_1,x_2)=
e^{-\ii\tfrac{\pi}{2}\sum_i s(O_i)}\,
\frac{\bra{0}\mathcal{T}_x\,O_1(\tau,x_1)\,
O_2(0,x_2)\ket{\mathrm{B}}}{\bra{0}\mathrm{B}\rangle}.
\end{equation}
Here $s(O_i)$ denotes the Lorentz spin of the operator $O_i$,
$\mathcal{T}_x$ is the $x$-ordering operator, which orders the largest
$x_i$ to the right, and $\ket{0}\in\mathcal{H}$ is the ground state of
the model on the infinite line.  The spin-dependent phase factor is
due to the rotation in Euclidean space; it was for example observed in
the Green function of the Ising model with a boundary magnetic
field~\cite{SE07}.  As we have interchanged space and time and $x$ is
running from $0$ to $-\infty$ in the new framework, the $\tau$- and
$x$-dependence of operators $O_i(\tau,x)$ is now given by
\begin{equation}
O_i(\tau,x)= e^{-xH}\,e^{-\ii\tau P}\,O_i(0,0)\,e^{\ii\tau P}\,e^{xH}, 
\label{eq:txdependence}
\end{equation}
where $H$ is the Hamiltonian of the system on the infinite line
$-\infty<\tau<\infty$, and $P$ is the total momentum.  

The boundary state, which encodes all informations on the boundary
condition, is given by
\begin{equation}
  \label{eq:boundarystate}
  \ket{\mathrm{B}}=\exp\left(\frac{1}{2}\int_{-\infty}^\infty\frac{d\xi}{2\pi}
    K^{ab}(\xi)Z_a^\dagger(-\xi)Z_b^\dagger(\xi)\right)\ket{0},
\end{equation}
where $K^{ab}(\xi)=R^b_{\bar{a}}(\text{i}\pi/2-\xi)$. For example, the
boundary reflection amplitudes $K$ stated in \eqref{eq:defK} and
\eqref{eq:KLE} are directly obtained from \eqref{eq:RLE} and
\eqref{eq:RSU2inv}.  For general $K_\mathrm{s}$ the amplitude $K^{ab}$
satisfies the boundary cross-unitarity
condition~\cite{GhoshalZamolodchikov94}
\begin{equation}
K^{ab}(\xi)=S^{ab}_{cd}(2\xi)\,K^{dc}(-\xi).
\label{eq:bcuc}
\end{equation}
Furthermore, for fixed boundary conditions we have $K^{\pm\pm}(\xi)=0$.

Below we calculate the spin part of the Green function
\eqref{eq:GFboson} using the boundary formalism presented
above. Specifically we will evaluate the correlation function
\eqref{eq:cfboundarystate}, where the operators $O_{1,2}$ are the
soliton-creating and -annihilating operators
$e^{\pm\frac{\ii}{2}\phi_\mathrm{s}}$ and
$e^{\pm\frac{\ii}{2}\bar{\phi}_\mathrm{s}}$ respectively. We define the
$n$-particle form factor of an arbitrary operator $O$ as
\begin{equation}
  f^O_{a_1,\ldots,a_n}(\theta_1,\ldots,\theta_n)=
  \bra{0}O\ket{\theta_1,\ldots,\theta_n}_{a_1,\ldots,a_n}=
  \bra{0}O\,Z^\dagger_{a_1}(\theta_1)\ldots Z^\dagger_{a_n}(\theta_n)\ket{0}.
  \label{eq:formfactordefinition}
\end{equation}
The form factors have to satisfy a set of relations, the so-called form-factor
axioms~\cite{Smirnov92book,Lukyanov95,Delfino04}, which we state for
completeness in App.~\ref{sec:ffaxioms}. As the operators
$e^{-\frac{\ii}{2}\phi_\mathrm{s}}$ and
$e^{\frac{\ii}{2}\bar{\phi}_\mathrm{s}}$ create one soliton, their respective
form factors \eqref{eq:formfactordefinition} vanish unless $\sum_i a_i=-1$.
The form factors containing up to three particles were derived by Lukyanov and
Zamolodchikov~\cite{LukyanovZamolodchikov01}.  In our conventions the
single-particle form factors are given by
\begin{equation}
\label{eq:ff1}
\bra{0}e^{-\frac{\ii}{2}\phi_\mathrm{s}}\ket{\theta}_-=
\sqrt{Z_1}\,e^{\text{i}\frac{\pi}{8}}\,e^{\theta/4},\quad
\bra{0}e^{\frac{\ii}{2}\bar{\phi}_\mathrm{s}}\ket{\theta}_-=
\sqrt{Z_1}\,e^{-\text{i}\frac{\pi}{8}}\,e^{-\theta/4},
\end{equation}
where the normalization constant $Z_1$ (not to be confused with the
Faddeev--Zamolodchikov operators $Z_\pm(\theta)$ and
$Z_\pm^\dagger(\theta)$) depends on $K_\mathrm{s}$ and can be found in
Ref.~[\onlinecite{LukyanovZamolodchikov01}]. Evaluation at the LEP
yields $Z_1\approx 3.32052\,\Delta^{5/8}$ whereas at the SU(2)
invariant point one finds $Z_1\approx 0.921862\,\Delta^{1/2}$.  The
three-particle form factors are known in terms of contour integrals,
which can be explicitly evaluated at the LEP:
\begin{equation}
\left.
\begin{array}{r}
\bra{0}e^{-\frac{\ii}{2}\phi_\mathrm{s}}
\ket{\theta_1,\theta_2,\theta_3}_{--+}\\[2mm]
\bra{0}e^{\frac{\ii}{2}\bar{\phi}_\mathrm{s}}
\ket{\theta_1,\theta_2,\theta_3}_{--+}
\end{array}\right\}
=-\text{i}\sqrt{\frac{Z_1}{2}}\,e^{\pm\text{i}\frac{\pi}{8}}\,
e^{\pm(\theta_1+\theta_2-\theta_3)/4}\,\frac{\sinh\frac{\theta_1-\theta_2}{2}}
{\cosh\frac{\theta_1-\theta_3}{2}\cosh\frac{\theta_2-\theta_3}{2}}.
\label{eq:ff3}
\end{equation}
The three-particle form factors for other orderings of the U(1)
indices can be easily obtained using the scattering axiom stated in
App.~\ref{sec:ffaxioms}.

The correlation functions to be calculated below contain matrix elements with
incoming and outgoing particles,
\begin{equation}
^{a_1,\ldots,a_n}\!\!\bra{\theta_1,\ldots,\theta_n}
O\ket{\xi_m,\ldots,\xi_1}_{b_m,\ldots,b_1},
\label{eq:generalmatrixelement}
\end{equation}
which possess kinematical poles whenever $\theta_i=\xi_j$ and
$a_i=b_j$.  These matrix elements can be decomposed into a
``connected'' and ``disconnected'' contributions. The latter are
characterized by the appearance of terms like
$\delta(\theta_i-\xi_j)$, signaling that some of the particles do not
encounter the operator $O$ in the process described by the matrix
element. We deal with these terms following ideas by
Smirnov~\cite{Smirnov92book} that allow us to analytically continue
form factors. Let $\overrightarrow{A}=\{\theta_1,\ldots,\theta_n\}$
with $\theta_1<\theta_2<\ldots<\theta_n$ and
$\overleftarrow{B}=\{\xi_m,\ldots,\xi_1\}$ with
$\xi_m>\xi_{m-1}>\ldots>\xi_1$ denote two sets of ordered rapidities and
introduce the notations
\begin{eqnarray}
Z[\overrightarrow{A}]_{a_1\ldots
  a_n}&\equiv& Z_{a_1}(\theta_1)Z_{a_2}(\theta_2)\ldots Z_{a_n}(\theta_n),\\*
Z^\dagger[\overleftarrow{B}]_{b_m\ldots b_1}
&\equiv&Z^\dagger_{b_m}(\xi_m) Z^\dagger_{b_{m-1}}(\xi_{m-1})\ldots
Z^\dagger_{b_1}(\xi_1).  
\end{eqnarray}
Now let $A_1$ and $A_2$ be a partition of $A$, i.e. $A=A_1\cup A_2$, where
$A_1$ contains $n(A_1)=n-k$ rapidities.  As a consequence of the
Faddeev--Zamolodchikov algebra we have
\begin{equation}
Z[\overrightarrow{A}]_{a_1\ldots
  a_n}=S(\overrightarrow{A}|\overrightarrow{A_1})^{c_1\ldots
  c_n}_{a_1\ldots a_n}\,Z[\overrightarrow{A_2}]_{c_1\ldots c_k}\,
Z[\overrightarrow{A_1}]_{c_{k+1}\ldots c_n},
\end{equation}
where $S(\overrightarrow{A}|\overrightarrow{A_1})$ is the product of
two-particle scattering matrices needed to rearrange the order of
Faddeev--Zamolodchikov operators in $Z[\overrightarrow{A}]$ to arrive
at $Z[\overrightarrow{A_2}]Z[\overrightarrow{A_1}]$.  For example, if
$\overrightarrow{A}=\{\theta_1,\ldots,\theta_4\}$ and
$\overrightarrow{A_1}=\{\theta_2,\theta_3\}$ it is given by
\begin{equation}
S(\overrightarrow{A}|\overrightarrow{A_1})^{c_1\ldots
c_4}_{a_1\ldots a_4}=\delta_{a_1}^{c_4}\,S_{a_2b}^{c_2c_4}(\theta_2-\theta_4)\,
S_{a_3a_4}^{c_3b}(\theta_3-\theta_4).
\end{equation}
Similarly we have
\begin{equation}
Z^\dagger[\overleftarrow{B}]_{b_m\ldots b_1}=
Z^\dagger[\overleftarrow{B_1}]_{d_m\ldots d_{l+1}}
Z^\dagger[\overleftarrow{B}_2]_{d_l\ldots d_1}
\ S(\overleftarrow{B_1}|\overleftarrow{B})_{b_m\ldots b_1}^{d_m\ldots d_1}.
\end{equation}
Finally we define
\begin{equation}
\delta[\overrightarrow{A},\overleftarrow{B}]_{a_1\ldots a_n
\atop b_m\ldots b_1}
=\delta_{nm}\prod_{j=1}^n
2\pi\delta_{a_jb_j}\delta(\theta_j-\xi_j).
\end{equation}
We are now in the position to analytically continue matrix elements as
\begin{eqnarray}
\langle 0|Z[\overrightarrow{A}]_{a_1\ldots a_n}\,O\,
Z^\dagger[\overleftarrow{B}]_{b_m\ldots b_1}|0\rangle&=&
\sum_{{A=A_1\cup A_2}\atop{B=B_1\cup B_2}}
S(\overrightarrow{A}|\overrightarrow{A_1})^{c_1\ldots c_n}_{a_1\ldots a_n}
\,S(\overleftarrow{B_1}|\overleftarrow{B})_{b_m\ldots b_1}^{d_m\ldots d_1}
\,\delta[\overrightarrow{A_2},\overleftarrow{B_2}]
_{c_1\ldots c_k\atop d_l\ldots d_1}\nonumber\\*
&&\qquad\qquad\times
\langle 0|Z[\overrightarrow{A}_1+\ii 0]_{c_{k+1}\ldots c_n}\,O\,
Z^\dagger[\overleftarrow{B}_1]_{d_m\ldots d_{l+1}}|0\rangle.
\label{eq:reg1}
\end{eqnarray}
Here the sum is over all possible ways to break the sets $A$ and $B$ into
subsets and $\overrightarrow{A}_1+\ii 0$ means that all rapidities in $A_1$
are slightly moved into the upper half-plane.  Similarly, we could choose to
analytically continue to the lower half-plane
\begin{eqnarray}
\langle 0|Z[\overrightarrow{A}]_{a_1\ldots a_n}\,O\,
Z^\dagger[\overleftarrow{B}]_{b_m\ldots b_1}|0\rangle&=&
\sum_{{A=A_1\cup A_2}\atop{B=B_1\cup B_2}}d_{A_2}(O)\,
S(\overrightarrow{A}|\overrightarrow{A_2})^{c_1\ldots c_n}_{a_1\ldots a_n}
\,S(\overleftarrow{B_2}|\overleftarrow{B})_{b_m\ldots b_1}^{d_m\ldots d_1}
\,\delta[\overrightarrow{A_2},\overleftarrow{B_2}]
_{c_1\ldots c_k\atop d_l\ldots d_1}\nonumber\\*
&&\qquad\qquad\times
\langle 0|Z[\overrightarrow{A}_1-\ii 0]_{c_{k+1}\ldots c_n}\,O\,
Z^\dagger[\overleftarrow{B}_1]_{d_m\ldots d_{l+1}}|0\rangle.
\label{eq:reg2}
\end{eqnarray}
The factor $d_{A_2}(O)$ is due to a possible semi-locality of the operator $O$
with respect to the fundamental fields creating the
excitations~\cite{YurovZamolodchikov91,Smirnov92book,Lukyanov95,Delfino04}.
If we use the operators $O^{\pm}_0$ defined in \eqref{eq:defOna} as
fundamental fields and denote the mutual semi-locality factor of $O$ and
$O^{\pm}_0$ by $l_\pm(O)$, it is given by
\begin{equation}
d_A(O)=\prod_{i=1}^{n(A)} l_{a_i}(O)\quad\Rightarrow\quad
d_{A_2}\Bigl(e^{-\frac{\ii}{2}\phi_\mathrm{s}}\Bigr)=
\prod_{i=1}^{k} e^{-\ii\frac{\pi}{2} a_i},\quad
d_{A_2}\Bigl(e^{\frac{\ii}{2}\bar{\phi}_\mathrm{s}}\Bigr)=
\prod_{i=1}^{k} e^{\ii\frac{\pi}{2} a_i}.
\end{equation}
The remaining matrix elements in \eqref{eq:reg1} and \eqref{eq:reg2} can be
evaluated using crossing
\begin{eqnarray}
&&\langle 0|Z[\overrightarrow{A}_1\pm\ii 0]_{c_{k+1}\ldots c_n}\,O\,
Z^\dagger[\overleftarrow{B}_1]_{d_m\ldots d_{l+1}}|0\rangle\,=\,
^{c_{k+1},\ldots,c_{n}}\!\bra{\theta_{i_{k+1}}\!\pm\!\ii 0,\ldots,
\theta_{i_n}\!\pm\!\ii 0}O
\ket{\xi_{j_m},\ldots,\xi_{j_{l+1}}}_{d_{m},\ldots,d_{l+1}}\nonumber\\*[2mm]
&&\qquad=d_{A_1}(O)\,C_{c_{k+1}e_{k+1}}\ldots C_{c_ne_n}\,
f^O_{e_{k+1},\ldots,e_{n},d_{m},\ldots,d_{l+1}}
(\theta_{i_{k+1}}\!+\!\ii\pi\!\pm\!\ii\eta_{i_{k+1}},\ldots,
\theta_{i_n}\!+\!\ii\pi\!\pm\!\ii\eta_{i_n},
\xi_{j_m},\ldots,\xi_{j_{l+1}}),\qquad\quad
\label{eq:crossing}
\end{eqnarray}
where $C_{ab}=\delta_{a+b,0}$ is the charge conjugation matrix and
$\eta_i\rightarrow 0+$. The analytic continuation of general matrix
elements \eqref{eq:generalmatrixelement} with arbitrary orders of the
rapidities can be obtained using the scattering axiom (see below).

\subsection{Form-factor axioms}\label{sec:ffaxioms}
For completeness we state here the used form-factor axioms. We follow
Delfino~\cite{Delfino04}. The $n$-particle form factor of an arbitrary
operator $O$ was defined in \eqref{eq:formfactordefinition}.  We use
the local bosonic fields
\begin{equation}
  O^{\pm}_0(\tau,x)=
  \exp\left(\mp\frac{1}{4K_\mathrm{s}}\int_{-\infty}^\tau\,d\tau\,
    \partial_x\,\Phi'_\mathrm{s}(\tau,x)\right),
  \label{eq:defOna}
\end{equation}
as fundamental fields for the creation of solitons and antisolitons.
The corresponding creation and annihilation operators are
$Z_\pm^\dagger(\theta)$ and $Z_\pm(\theta)$ introduced in
\eqref{eq:defA}. The form-factor axioms read:
\begin{enumerate}
\item The form factors $f^O_{a_1,\ldots,a_n}(\theta_1,\ldots,\theta_{n})$
  are meromorphic functions of $\theta_n$ in the physical strip $0\le
  \mathfrak{Im}\,\theta_n\le 2\pi$. There exist only simple poles in this
  strip.
\item Scattering axiom:
\begin{equation*}
\begin{split}
&f^O_{a_1,\ldots,a_{i},a_{i+1},\ldots,a_n}
(\theta_1,\ldots,\theta_{i},\theta_{i+1},\ldots,\theta_n)\\
&\hspace{10mm}
=S_{a_{i}a_{i+1}}^{b_ib_{i+1}}(\theta_{i}-\theta_{i+1})\,
f^O_{a_1,\ldots,b_{i+1},b_{i},\ldots,a_n}
(\theta_1,\ldots,\theta_{i+1},\theta_{i},\ldots,\theta_n),
\end{split}
\end{equation*}
with the scattering matrix
$S^{b_{i}b_{i+1}}_{a_ia_{i+1}}(\theta_{i}-\theta_{i+1})$. At the free-fermion
point it is given by
$S_{a_1a_2}^{b_1b_2}(\theta)=-\delta_{a_1}^{b_1}\delta_{a_2}^{b_2}$.
\item Periodicity axiom:
\begin{equation*}
f^O_{a_1,\ldots,a_n}(\theta_1+2\pi\ii,\theta_2,\ldots,\theta_n)=
l_{a_1}(O)\,
f^O_{a_2,\ldots,a_n,a_1}(\theta_2,\ldots,\theta_{n},\theta_1),
\end{equation*}
where $l_{\pm}(O)$ is the mutual semi-locality factor between the
operator $O$ and the fundamental fields $O_0^{\pm}$. In particular, we
have $l_\pm(e^{-\frac{\ii}{2}\phi_\mathrm{s}})=\mp\ii$ and
$l_\pm(e^{\frac{\ii}{2}\bar{\phi}_\mathrm{s}})=\pm\ii$.
\item Lorentz transformations:
\begin{equation*}
f^O_{a_1,\ldots,a_n}(\theta_1+\Lambda,\ldots,\theta_n+\Lambda)=
e^{s(O)\Lambda}\,f^O_{a_1,\ldots,a_n}(\theta_1,\ldots,\theta_n),
\end{equation*}
where $s(O)$ denotes the Lorentz spin of $O$. Here we have
$s(e^{\pm\frac{\ii}{2}\phi_\mathrm{s}})=1/4$ and
$s(e^{\pm\frac{\ii}{2}\bar{\phi}_\mathrm{s}})=-1/4$.
\item Annihilation pole axiom:
\begin{equation*}
\begin{split}
&\mathrm{Res}\bigl[
f^O_{a,b,a_1,\ldots,a_n}(\theta',\theta,\theta_1,\ldots,\theta_n),
\theta'=\theta+\ii\pi\bigr]\\
&\hspace{5mm}=\ii\,C_{ac}\,
f^O_{b_1,\ldots,b_n}(\theta_1,\ldots,\theta_n)
\left[\delta_{a_1}^{b_1}\ldots\delta_{a_n}^{b_n}\delta_{b}^{c}-
l_a(O) S_{b_{\phantom{1}}a_1}^{c_1b_1}(\theta-\theta_1)
S_{c_1a_2}^{c_2b_2}(\theta-\theta_2)
\ldots
S_{c_{n-1}a_n}^{c_{\phantom{n-1}}b_n}(\theta-\theta_n)\right],
\end{split}
\end{equation*}
with the charge conjugation matrix $C_{ab}=\delta_{a+b,0}$.  If there do not
exist bound states in the model, i.e. for $K_\mathrm{s}^2\ge 1/2$, these
are the only poles of the form factors.
\end{enumerate}
We note that the precise form of the axioms depends on the basis of
scattering states and thus changes under a unitary transformation of
the operators $Z_a(\theta)$.

\subsection{Correlation functions}
In this appendix we derive \eqref{eq:GRLspin} using the boundary state
formalism. We start with the spin part of \eqref{eq:GFboson}. After the
rotation in Euclidean space this is given by \eqref{eq:cfboundarystate}. We
insert a resolution of the identity \eqref{eq:residentity} and expand the
boundary state \eqref{eq:boundarystate} in powers of $K$. This yields the
double expansion ($\tau>0$, $x_1<x_2$)
\begin{equation}
\label{eq:corr}
\Big\langle e^{-\frac{\ii}{2}f_\sigma\phi_\mathrm{s}(\tau,x_1)}\,
e^{-\frac{\ii}{2}f_{\sigma'}\bar{\phi}_\mathrm{s}(0,x_2)}\Big\rangle_\mathrm{s}
=\bra{0} e^{-\frac{\ii}{2}f_\sigma\phi_\mathrm{s}(\tau,x_1)}\,
e^{-\frac{\ii}{2}f_{\sigma'}\bar{\phi}_\mathrm{s}(0,x_2)}\ket{\mathrm{B}}
=\delta_{\sigma\sigma'}
\sum_{n=0}^\infty\sum_{m=0}^\infty C_{n\,2m}(\tau,x_1,x_2),
\end{equation}
where we have used $s(e^{\pm\frac{\ii}{2}\phi_\mathrm{s}})+
s(e^{\pm\frac{\ii}{2}\bar{\phi}_\mathrm{s}})=0$. The operators
$e^{\pm\frac{\ii}{2}\phi_\mathrm{s}}$ and
$e^{\pm\frac{\ii}{2}\bar{\phi}_\mathrm{s}}$ change the U(1) charge by $\mp 1$
and $\pm 1$, respectively.  As the boundary state has vanishing U(1) charge
for Dirichlet boundary conditions ($K^{\sigma\sigma}(\xi)=0$) the correlation
function is diagonal in spin space. Furthermore we have defined the auxiliary
functions
\begin{equation}
\begin{split}
C_{n\,2m}(\tau,x_1,x_2)=&\frac{1}{2^m}\frac{1}{m!}\frac{1}{n!}
\int_{-\infty}^\infty\frac{d\xi_1\ldots d\xi_m}{(2\pi)^m}
\int_{-\infty}^\infty\frac{d\theta_1\ldots d\theta_{n}}{(2\pi)^{n}}\,
K^{a_1b_1}(\xi_1)\ldots K^{a_mb_m}(\xi_m)\label{eq:corrmn}\\*[2mm]
&\hspace{-20mm}\times
\bra{0}e^{-\frac{\ii}{2}f_\sigma\phi_\mathrm{s}(\tau,x_1)}
\ket{\theta_{n},\ldots,\theta_{1}}_{c_n,\ldots,c_1}\;
^{c_1,\ldots,c_n}\!\bra{\theta_{1},\ldots,\theta_{n}}
e^{-\frac{\ii}{2}f_\sigma\bar{\phi}_\mathrm{s}(0,x_2)}
\ket{-\xi_1,\xi_1,\ldots,-\xi_m,\xi_m}_{a_1,b_1,\ldots,a_m,b_m}.
\end{split}
\end{equation}
We use the notations $\up=+$, $\dw=-$, $\bar{\sigma}=-$ for $\sigma=+$ and
vice versa.  We label the various terms in the expansion (\ref{eq:corr}) by
the numbers of particles in the intermediate state $n$ and in the boundary
state $2m$, respectively. The $\tau$- and $x$-dependence of the operators is
given by \eqref{eq:txdependence}. We have already assumed $x_1<x_2$ to avoid
additional phases due to the mutual semi-locality of the operators. For the
calculation of the LDOS we have to take $x_1\rightarrow x_2-$ finally. The
second matrix element possesses kinematical poles which we treat using
\eqref{eq:reg1}. This introduces a third, finite summation in \eqref{eq:corr},
which labels the ``connectedness'' of the corresponding terms. We note,
however, that \eqref{eq:reg1} and \eqref{eq:reg2} yield the same results.

Let us start with the first non-vanishing term in the series \eqref{eq:corr},
which is using \eqref{eq:txdependence} given by (we recall that the
center-of-mass coordinates are defined by $R=(x_1+x_2)/2<0$ and $r=x_1-x_2<0$)
\begin{equation}
\label{eq:C10-1}
C_{10}=\int_{-\infty}^\infty\frac{d\theta}{2\pi}\,
\bra{0}e^{-\frac{\ii}{2}f_\sigma\phi_\mathrm{s}}\ket{\theta}_{c}\;
^{c}\!\bra{\theta}e^{-\frac{\ii}{2}f_\sigma\bar{\phi}_\mathrm{s}}\ket{0}\,
e^{\frac{\Delta}{v_\mathrm{s}}r\cosh\theta}\,e^{\ii\Delta\tau\sinh\theta}
=Z_1\,e^{\ii\frac{\pi}{4}}\int_{-\infty}^\infty\frac{d\theta}{2\pi}\,
e^{\frac{\Delta}{v_\mathrm{s}}r\cosh\theta}\,e^{\ii\Delta\tau\sinh\theta}.
\end{equation}
We can rewrite this by shifting the contour of integration as
$\theta\rightarrow\theta+\text{i}\pi/2$. The contributions of
$\mathfrak{Re}\,\theta=\pm\infty$ vanish due to the exponential factors. As
there are no poles in the strip $0\le\mathfrak{Im}\,\theta\le\pi/2$ we
find
\begin{equation}
\label{eq:C10-2}
C_{10}=Z_1\,e^{\ii\frac{\pi}{4}}\int_{-\infty}^\infty\frac{d\theta}{2\pi}\,
e^{\ii\frac{\Delta}{v_\mathrm{s}}r\sinh\theta}\,e^{-\Delta\tau\cosh\theta}=
\frac{Z_1}{\pi}\,e^{\ii\frac{\pi}{4}}\,
K_0\bigl(\Delta\sqrt{\tau^2+r^2/v_\mathrm{s}^2}\bigr),
\end{equation} 
where $K_0$ denotes the modified Bessel function~\cite{GradshteynRyzhik80}.

The first term containing the boundary reflection amplitude $K$ is
$C_{12}$. For $f_\sigma=-1$ it reads
\begin{equation}
C_{12}=\frac{1}{2}\int_{-\infty}^\infty\frac{d\xi}{2\pi}\frac{d\theta}{2\pi}\,
K^{ab}(\xi)\,\bra{0}e^{\frac{\ii}{2}\phi_\mathrm{s}}\ket{\theta}_{c}\;
^{c}\!\bra{\theta}e^{\frac{\ii}{2}\bar{\phi}_\mathrm{s}}\ket{-\xi,\xi}_{ab}\,
e^{\frac{\Delta}{v_\mathrm{s}}r\cosh\theta}\,
e^{2{\frac{\Delta}{v_\mathrm{s}}x_2\cosh\xi}}\,e^{\ii\Delta\tau\sinh\theta},
\end{equation}
The first form factor vanishes for $c\neq +$ and can be evaluated using
\eqref{eq:crossing}
\begin{equation}
\bra{0}e^{\frac{\ii}{2}\phi_\mathrm{s}}\ket{\theta}_{+}=
^+\!\bra{\theta}e^{-\frac{\ii}{2}\phi_\mathrm{s}}\ket{0}^*=
e^{\ii\frac{\pi}{2}}\,
\bra{0}e^{-\frac{\ii}{2}\phi_\mathrm{s}}\ket{\theta+\ii\pi}_-^*=
\sqrt{Z_1}\,e^{\ii\frac{\pi}{8}}\,e^{\theta/4}.
\label{eq:C12a}
\end{equation}
For the second matrix element we use \eqref{eq:reg1}, which explicitly yields
\begin{equation}
\begin{split}
&^+\!\bra{\theta}e^{\frac{\ii}{2}\bar{\phi}_\mathrm{s}}\ket{-\xi,\xi}_{-+}=
^+\!\bra{\theta+\ii 0}e^{\frac{\ii}{2}\bar{\phi}_\mathrm{s}}\ket{-\xi,\xi}_{-+}
+2\pi\delta(\theta-\xi)\,
\bra{0}e^{\frac{\ii}{2}\bar{\phi}_\mathrm{s}}\ket{-\xi}_-
+2\pi\delta(\theta+\xi)\,S_{-+}^{+-}(-2\xi)\,
\bra{0}e^{\frac{\ii}{2}\bar{\phi}_\mathrm{s}}\ket{\xi}_-,\\*
&^+\!\bra{\theta}e^{\frac{\ii}{2}\bar{\phi}_\mathrm{s}}\ket{-\xi,\xi}_{+-}=
^+\!\bra{\theta+\ii 0}e^{\frac{\ii}{2}\bar{\phi}_\mathrm{s}}\ket{-\xi,\xi}_{+-}
+2\pi\delta(\theta+\xi)\,S_{+-}^{+-}(-2\xi)\,
\bra{0}e^{\frac{\ii}{2}\bar{\phi}_\mathrm{s}}\ket{\xi}_-.
\end{split}
\label{eq:C12reg}
\end{equation}
This leads to two contributions which we denote by $C_{12}^0$ and $C_{12}^1$
respectively. The additional upper index denotes the number of lines
connecting the operators (the ``connectedness''), i.e. the number of internal
$\theta$-integrations left after using \eqref{eq:reg1}. The first terms on the
right-hand side of \eqref{eq:C12reg} in each equation together yield
$C_{12}^1$. We will calculate this term at the LEP in the next section. On the
other hand, the disconnected piece is given by
\begin{equation}
C_{12}^0=\frac{Z_1}{2}\int_{-\infty}^\infty\frac{d\xi}{2\pi}\,
\Bigl[K^{-+}(\xi)+K^{+-}(-\xi)\,S^{+-}_{+-}(2\xi)+
K^{-+}(-\xi)\,S^{+-}_{-+}(2\xi)\Bigr]\,e^{\xi/2}\,
e^{2{\frac{\Delta}{v_\mathrm{s}}R\cosh\xi}}\,e^{\ii\Delta\tau\sinh\xi}.
\label{eq:C121}
\end{equation}
Using the boundary cross-unitarity \eqref{eq:bcuc} the terms in the square
brackets equal $2K^{-+}(\xi)$. With the similar calculation for $f_\sigma=1$
we arrive at
\begin{equation}
C_{12}^0=Z_1\int_{-\infty}^\infty\frac{d\xi}{2\pi}\,
K^{\sigma\bar{\sigma}}(\xi)\,e^{\xi/2}\,
e^{2{\frac{\Delta}{v_\mathrm{s}}R\cosh\xi}}\,e^{\ii\Delta\tau\sinh\xi},
\label{eq:C120result}
\end{equation}
The second term in \eqref{eq:GRLspin} is now obtained by shifting the contour
of integration as $\xi\rightarrow\xi+\ii\pi/2$ while noting that for the
boundary condition $\Phi_\mathrm{s}(x=0)=0$ the reflection amplitude does not
depend on $\sigma$ and is analytic in the physical strip
$0\le\mathfrak{Im}\,\xi\le\pi/2$. If $\Phi_\mathrm{s}(x=0)\neq 0$, however,
the reflection amplitude may possess a pole in the physical strip.  We will
calculate the resulting term in App.~\ref{sec:appbbs}.

\subsection{Higher-order terms}\label{sec:appho}
In order to estimate the truncation error in \eqref{eq:GRLspin}, we
will calculate the leading corrections due to a higher number of
particles in the intermediate state as well as higher-order
corrections due to the boundary.  The resulting corrections to the
LDOS are discussed in Sec.~\ref{sec:LDOShigherorders}. We will
restrict ourselves to the LEP, where the form factors are given by
\eqref{eq:ff3},
$S_{a_1a_2}^{b_1b_2}(\theta)=-\delta_{a_1}^{b_1}\delta_{a_2}^{b_2}$,
and $K^{ab}(\xi)=-K^{ba}(-\xi)$.

The leading correction due to a higher number of particles in the intermediate
state is given by $C_{30}$, 
\begin{equation}
C_{30}=Z_1\,\frac{e^{\ii\frac{\pi}{4}}}{4}\int_{-\infty}^\infty
\frac{d\theta_1d\theta_2d\theta_3}{(2\pi)^3}\,
\frac{\sinh^2\frac{\theta_1-\theta_2}{2}}
{\cosh^2\frac{\theta_1-\theta_3}{2}\,\cosh^2\frac{\theta_2-\theta_3}{2}}\,
e^{\ii\frac{\Delta}{v_\mathrm{s}}r\sum_i\sinh\theta_i}\,
e^{-\Delta\tau\sum_i\cosh\theta_i}.
\end{equation}
The resulting contribution to the LDOS discussed in
Sec.~\ref{sec:LDOShigherorders} is denoted by $N_{30}$. We note that
$C_{20}=0$.

The first sub-leading term due to the boundary is given by $C_{12}^1$, i.e.
the connected piece of $C_{12}$ obtained from the first terms in
\eqref{eq:C12reg}. For $f_\sigma=-1$ this term reads using \eqref{eq:C12a} and
\eqref{eq:ff3}
\begin{equation}
C_{12}^1=Z_1\,\frac{e^{-\ii\frac{\pi}{4}}}{\sqrt{2}}\int_{-\infty}^\infty
\frac{d\xi}{2\pi}\frac{d\theta}{2\pi}\,\frac{K^{-+}(\xi)}{\cosh\xi}\,
\frac{\sinh\frac{\theta+\xi+\ii\pi}{2}}
{\cosh\frac{\theta-\xi+\ii\pi+\ii\eta}{2}}\,
e^{\xi/2}\,e^{\frac{\Delta}{v_\mathrm{s}}r\cosh\theta}\,
e^{2\frac{\Delta}{v_\mathrm{s}}x_2\cosh\xi}\,
e^{\ii\Delta\tau\sinh\theta},
\end{equation}
where $\eta\rightarrow 0+$. We can handle the singularity at
$\theta=\xi-\ii\eta$ by shifting
$\theta\rightarrow\theta+\ii\pi/2$. Performing the same steps for $f_\sigma=1$
we arrive at
\begin{equation}
C_{12}^1=Z_1\,\frac{e^{-\ii\frac{\pi}{4}}}{\sqrt{2}}\int_{-\infty}^\infty
\frac{d\xi}{2\pi}\frac{d\theta}{2\pi}\,
\frac{K^{\sigma\bar{\sigma}}(\xi)}{\cosh\xi}\,
\frac{e^{\xi+\theta}-\ii}{\ii e^\xi+e^\theta}\,
e^{\xi/2}\,e^{\ii\frac{\Delta}{v_\mathrm{s}}r\sinh\theta}\,
e^{2\frac{\Delta}{v_\mathrm{s}}x_2\cosh\xi}\,
e^{-\Delta\tau\cosh\theta}.
\end{equation}
The next term in the series \eqref{eq:corr} is $C_{32}$, its disconnected
piece is similar to $C_{12}^1$ and reads explicitly
\begin{equation}
C_{32}^1=-Z_1\,\frac{e^{-\ii\frac{\pi}{4}}}{\sqrt{2}}\int_{-\infty}^\infty
\frac{d\xi}{2\pi}\frac{d\theta}{2\pi}\,
\frac{K^{\sigma\bar{\sigma}}(\xi)}{\cosh\xi}\,
\frac{\ii e^\xi+e^\theta}{e^{\xi+\theta}-\ii}\,
e^{\xi/2}\,e^{\ii\frac{\Delta}{v_\mathrm{s}}r\sinh\theta}\,
e^{2\frac{\Delta}{v_\mathrm{s}}x_1\cosh\xi}\,
e^{-\Delta\tau\cosh\theta}.
\end{equation}
The terms $C_{12}^1$ and $C_{32}^1$ are of the same order and yield together
the contribution to the LDOS denoted by $N_{11}$.

The term resulting in $N_{21}$ is 
\begin{eqnarray}
C_{32}^2&=&-\frac{Z_1}{2}\,e^{\ii\frac{\pi}{4}}
\int_{-\infty}^\infty\frac{d\xi}{2\pi}
\frac{d\theta_1d\theta_2}{(2\pi)^2}\,
e^{\ii\frac{\Delta}{v_\mathrm{s}}r\sum_i\sinh\theta_i}\,
e^{2\ii\frac{\Delta}{v_\mathrm{s}}R\sinh\xi}\,
e^{-\Delta\tau(\sum_i\cosh\theta_i+\cosh\xi)}\nonumber\\*
&&\hspace{20mm}\times
\left[\frac{K^{\sigma\bar{\sigma}}\bigl(\xi+\ii\tfrac{\pi}{2}\bigr)
\,e^{\xi/2}}{\cosh^2\frac{\theta_1-\theta_2}{2}}
\frac{\sinh\frac{\xi-\theta_1}{2}\,\sinh\frac{\xi+\theta_1}{2}}
{\cosh\frac{\xi-\theta_2}{2}\,\cosh\frac{\xi+\theta_2}{2}}-
\frac{\ii}{2}
\frac{K^{\bar{\sigma}\sigma}\bigl(\xi+\ii\tfrac{\pi}{2}\bigr)
\,e^{-\xi/2}\,\sinh^2\frac{\theta_1-\theta_2}{2}}
{\prod_i\cosh\frac{\xi-\theta_i}{2}\,\cosh\frac{\xi+\theta_i}{2}}\right].
\end{eqnarray}
We note that after analytic continuation $\tau\rightarrow\ii t+\delta$
and Fourier transformation $t\rightarrow E$ the exponential factor
$e^{-\Delta\tau(\sum_i\cosh\theta_i+\cosh\xi)}$ results in a vanishing
of $N_{21}$ for energies $E<3\Delta$.

The final term we wish to evaluate explicitly is the disconnected piece of
$C_{14}$. Considering first $f_\sigma=-1$ and keeping in mind that we have
restricted ourselves to the LEP, we can start with
\begin{eqnarray}
C_{14}&=&\frac{1}{2}\int_{-\infty}^\infty\frac{d\xi_1 d\xi_2}{(2\pi)^2}
\frac{d\theta}{2\pi}\,K^{-+}(\xi_1)\,K^{-+}(\xi_2)\,
e^{\frac{\Delta}{v_\mathrm{s}}r\cosh\theta}\,
e^{2\frac{\Delta}{v_\mathrm{s}}x_2\sum_i\cosh\xi_i}\,
e^{\ii\Delta\tau\sinh\theta}\nonumber\\*
&&\hspace{20mm}\times\bra{0}e^{\frac{\ii}{2}\phi_\mathrm{s}}\ket{\theta}_+\;
^+\!\bra{\theta}e^{\frac{\ii}{2}\bar{\phi}_\mathrm{s}}
\ket{-\xi_1,\xi_1,-\xi_2,\xi_2}_{-+-+}.
\end{eqnarray}
In the second matrix element we keep only the disconnected piece
\begin{equation}
^+\!\bra{\theta}e^{\frac{\ii}{2}\bar{\phi}_\mathrm{s}}
\ket{-\xi_1,\xi_1,-\xi_2,\xi_2}_{-+-+}=
2\pi\delta(\theta-\xi_1)\,\bra{0}e^{\frac{\ii}{2}\bar{\phi}_\mathrm{s}}
\ket{-\xi_1,-\xi_2,\xi_2}_{--+}+
2\pi\delta(\theta-\xi_2)\,\bra{0}e^{\frac{\ii}{2}\bar{\phi}_\mathrm{s}}
\ket{-\xi_1,\xi_1,-\xi_2}_{-+-}+\ldots
\end{equation}
In the resulting term $C_{14}^0$ we can shift the contour of integration,
$\xi_1\rightarrow\xi_1+\ii\pi/2$, to obtain
\begin{equation}
C_{14}^0=-\frac{Z_1}{\sqrt{2}}\,e^{-\ii\frac{\pi}{4}}\int_{-\infty}^\infty
\frac{d\xi_1 d\xi_2}{(2\pi)^2}\,
K^{\sigma\bar{\sigma}}\bigl(\xi_1+\ii\tfrac{\pi}{2}\bigr)\,
K^{\sigma\bar{\sigma}}(\xi_2)\,\frac{e^{(\xi_1+\xi_2)/2}}{\cosh\xi_2}\,
\frac{e^{\xi_1}+\ii e^{\xi_2}}{e^{\xi_1+\xi_2}-\ii}\,
e^{2\ii\frac{\Delta}{v_\mathrm{s}}R\sinh\xi_1}\,
e^{2\frac{\Delta}{v_\mathrm{s}}x_2\cosh\xi_2}\,
e^{-\Delta\tau\cosh\xi_1}.
\end{equation}
In the last step we have assumed that there exist no boundary bound
states (see below).  Finally, we mention that the next term,
$C_{34}^0$, equals $C_{14}^0$ with the coordinates $x_1$ and $x_2$
interchanged. These two terms together yield $N_{02}$.

The remaining two terms $N_{12}$ and $N_{03}$ discussed in
Sec.~\ref{sec:LDOShigherorders} follow from
$C_{14}^1+C_{34}^1+C_{54}^1$ and $C_{16}^0+C_{36}^0+C_{56}^0$
respectively.

\subsection{Boundary bound states}\label{sec:appbbs}
As discussed above, general Dirichlet boundary conditions
$\Phi_\mathrm{s}(0)=\Phi_\mathrm{s}^0\neq 0$ can result in the
appearance of a boundary bound state.  If $K_\mathrm{s}^2 \pi <
\Phi_\mathrm{s}^0<\pi$ the boundary reflection amplitude $K^{-+}(\xi)$
has a pole in the physical strip $0\le\mathfrak{Im}\,\xi\le\pi/2$
located at~\cite{SkorikSaleur95,MattssonDorey00} $\xi=\ii
(\pi-\Phi_\mathrm{s}^0)/(2-2K_\mathrm{s}^2)$. On the other hand,
$K^{+-}(\xi)$ is analytic in the physical strip but has a pole for
$-\pi/2\le\mathfrak{Im}\,\xi\le 0$.  We write the respective residues
as
\begin{equation}
\ii\,\text{Res}\Bigl[K^{\mp\pm}(\xi),\xi=\pm\ii\,\gamma\Bigr]=B\ge 0,
\qquad \gamma=\frac{\pi-\Phi_\mathrm{s}^0}{2-2K_\mathrm{s}^2},
\label{constantB}
\end{equation}
where $B$ depends on $K_\mathrm{s}$ only. We have checked the sign of $B$ by
performing an explicit mode expansion at the LEP as well as studying the
spectral function of the correlator $\bra{0_\mathrm{b}}e^{-\ii a
  \Phi'_\mathrm{s}(\tau,R)}\, e^{\ii a \Phi'_\mathrm{s}(0,R)}
\ket{0_\mathrm{b}}$ (for which the relevant form factors were obtained in
Ref.~[\onlinecite{Lukyanov97}]).

In the presence of a boundary bound state the poles of $K^{\mp\pm}(\xi)$ will
contribute whenever we shift the contour of integration
$\xi\rightarrow\xi\pm\ii\pi/2$ in a given term in the form-factor expansion
\eqref{eq:corr}. The leading term of this type is obtained from
\eqref{eq:C120result}, which yields \eqref{eq:GRLbbs} by a straightforward
calculation. The sub-leading term can be obtained similarly from $C_{14}^0$
and $C_{34}^0$. At the LEP it is given by ($B=-2\cos\Phi_\mathrm{s}^0$,
$\pi/2\le\Phi_\mathrm{s}^0\le\pi$)
\begin{equation}
\Theta\Bigl(\Phi_\mathrm{s}^0-\tfrac{\pi}{2}\Bigr)\,
\delta_{\sigma\dw}\,Z_1\sqrt{8}\,e^{-\frac{\ii}{2}\Phi_\mathrm{s}^0}
\cos\Phi_\mathrm{s}^0\,
e^{-2\frac{\Delta}{v_\mathrm{s}}R\cos\Phi_\mathrm{s}^0}\,
e^{-\Delta\tau\sin\Phi_\mathrm{s}^0}
\int_{-\infty}^\infty\frac{d\xi}{2\pi}\,
\frac{K^{\sigma\bar{\sigma}}(\xi)}{\cosh\xi}\,
\frac{\cosh\left(\tfrac{\xi}{2}+\tfrac{\ii}{2}\Phi_\mathrm{s}^0\right)}
{\sinh\left(\tfrac{\xi}{2}-\tfrac{\ii}{2}\Phi_\mathrm{s}^0\right)}\,
e^{\xi/2}\,e^{2\frac{\Delta}{v_\mathrm{s}}x_2\cosh\xi}.
\end{equation}

\section{Fourier transformation of the LDOS}\label{sec:FT}
We calculate the auxiliary function
\begin{eqnarray}
I(\omega,k)&=&\int_{-\infty}^0 dR\int_{-\infty}^\infty dt\,
\frac{e^{\ii(\omega t-kR)}}{\bigl(v_\mathrm{c}\tau-2\ii R\bigr)^a}\,
\frac{1}{\bigl(v_\mathrm{c}\tau+2\ii R\bigr)^b}\,
\left(\frac{2R}{v_\mathrm{c}\tau}\right)^{2c}
\Bigg|_{\tau\rightarrow\ii t+\delta}\\*
&=&\int_{-\infty}^0 dR\int_{-\infty}^\infty dt\,
\frac{e^{\ii(\omega t-kR)}}{\bigl(v_\mathrm{c}t-2R-\ii\delta\bigr)^a}\,
\frac{(-\ii)^{a+b+2c}}{\bigl(v_\mathrm{c}t+2R-\ii\delta\bigr)^b}\,
\left(\frac{2R}{v_\mathrm{c}t-\ii\delta}\right)^{2c},
\end{eqnarray}
where $v_\mathrm{c},a,b,c\in\mathbb{R}$, $v_\mathrm{c}>0$, $a+b<2$ and
$c>-1/2$.  We substitute $R\rightarrow-R$ and $t\rightarrow-t$, introduce
$s=v_\mathrm{c}t/2R$ and $\eta\rightarrow 0+$, and perform the resulting
$R$-integral (3.381.4 in Ref.~[\onlinecite{GradshteynRyzhik80}]), which yields
\begin{equation}
I(\omega,k)=
-\frac{e^{\ii\pi(a+b-c)}\,\Gamma(2-a-b)}{2^{a+b-1}\,v_\mathrm{c}}
\int_{-\infty}^\infty ds\,
\frac{\bigl(\tfrac{2\omega}{v_\mathrm{c}}s-k-\ii\eta\bigr)^{a+b-2}}
{(s-1+\ii\delta)^a\,(s+1+\ii\delta)^b\,(s+\ii\delta)^{2c}}.
\end{equation}
For $\omega<0$ the integrand has all its branch points in the lower half plane
and the integral over $s$ vanishes as long as $c>-1/2$. Hence we find
($s_0=v_\mathrm{c}k/2\omega$)
\begin{equation}
I(\omega,k)=-\Theta(\omega)\,\Gamma(2-a-b)
\frac{e^{\ii\pi(a+b-c)}\,\omega^{a+b-2}}{2\,v_\mathrm{c}^{a+b-1}}
\int_{-\infty}^\infty ds\,
\frac{\bigl(s-s_0-\ii\eta\bigr)^{a+b-2}}
{(s-1+\ii\delta)^a\,(s+1+\ii\delta)^b\,(s+\ii\delta)^{2c}}.
\label{eq:ft1}
\end{equation}

\begin{figure}[t]
\centering
\includegraphics[scale=0.2]{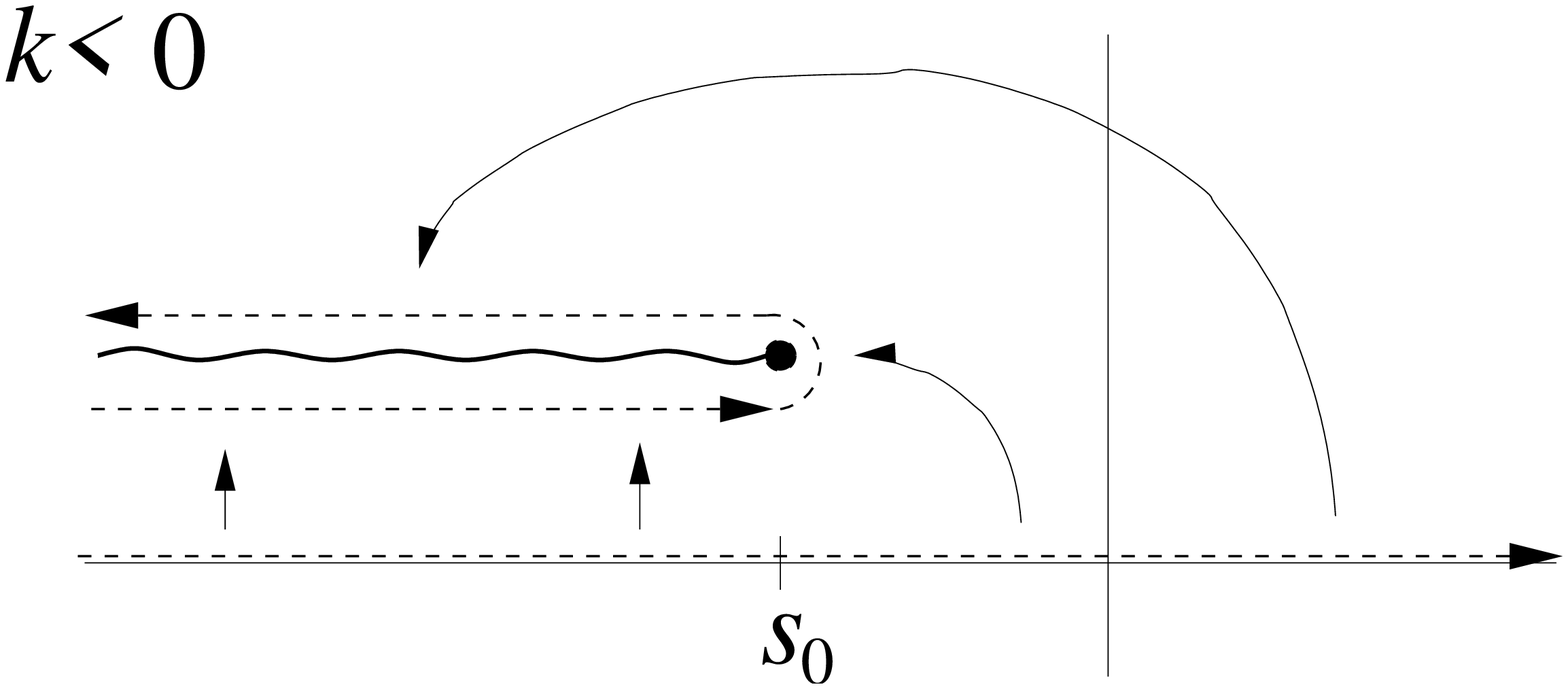}\hspace{10mm}
\includegraphics[scale=0.2]{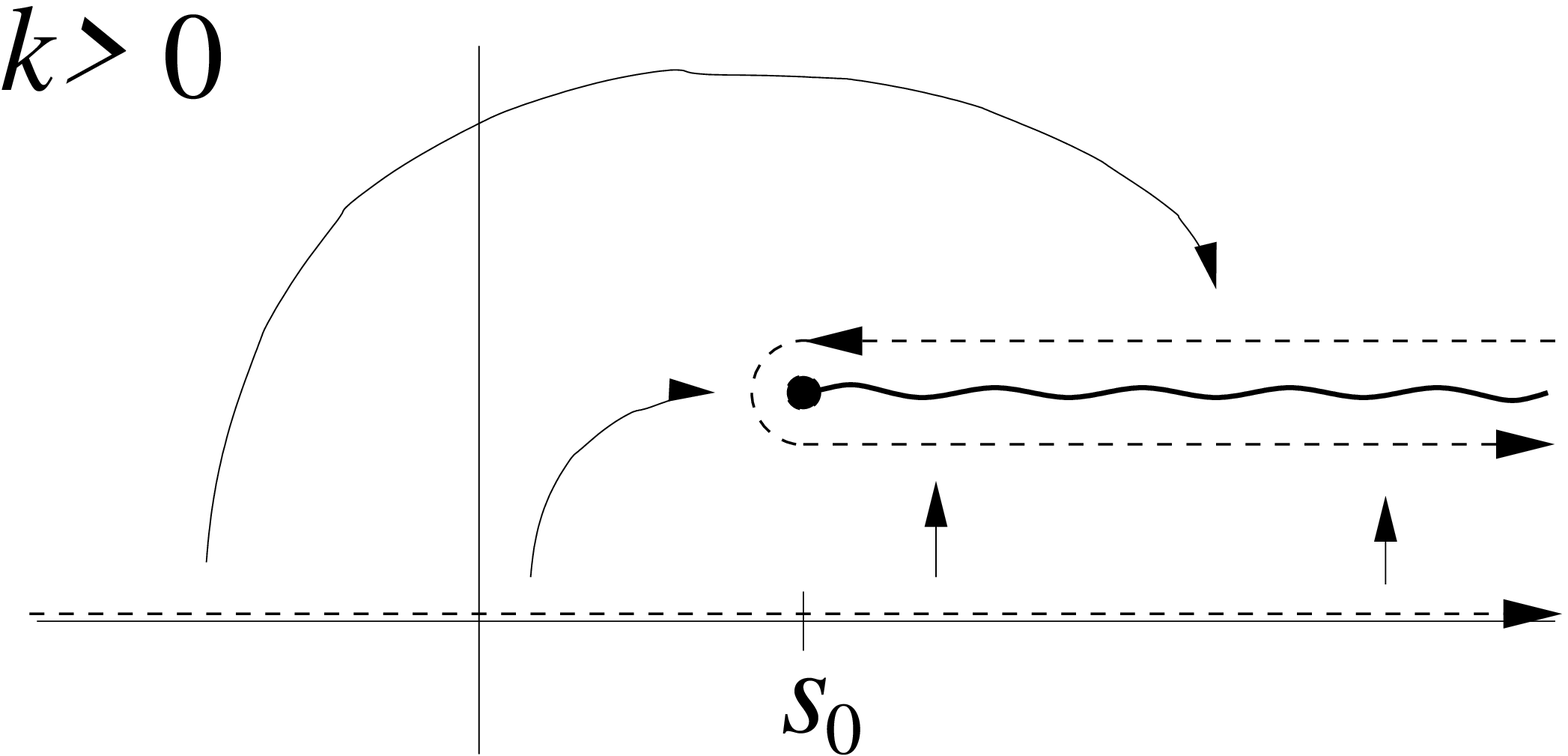}  
\caption{Branch cut and deformation of the contour of integration used 
for $k<0$ and $k>0$ respectively.}
\label{fig:contour}
\end{figure}
First consider the case $k<0$. The numerator of the integrand has a branch
point at $s=s_0+\ii\eta$ in the upper half plane. We place the cut running
from $-\infty+\ii\eta$ to $s_0+\ii\eta$ with constant imaginary part (see
Fig.~\ref{fig:contour}).  Now we deform the contour of integration and rewrite
the integration above the cut as integration below the cut using
\begin{equation}
\int_{-\infty}^\infty ds\,(s-s_0-\ii\eta)^{a+b-2}\,g(s)=
\int_{-\infty}^{s_0} ds\,(s-s_0-\ii\eta)^{a+b-2}\,g(s)\,
\Bigl[1-e^{2\pi\ii (a+b)}\Bigr].
\label{eq:cut}
\end{equation}
Assuming $1<a+b$ and substituting $s=s_0/t$ this yields for the integral in
\eqref{eq:ft1}
\begin{equation}
-2\ii\,\sin\bigl(\pi(a+b)\bigr)\,|s_0|^{a+b-1}\,
\int_{0}^1 dt\,\frac{t^{2c}\,(1-t)^{a+b-2}}
{(s_0-t+\ii\delta)^a\,(s_0+t+\ii\delta)^b\,(s_0+\ii\delta)^{2c}}.
\end{equation}
Finally, using (recall $s_0<0$, $\delta=0+$)
\begin{eqnarray}
\bigl(s_0-t+\ii\delta\bigr)^{-a}&=&
\bigl(s_0(1-(1/s_0+\ii\delta)t)\bigr)^{-a}=
|s_0|^{-a}\,e^{-\ii\pi a}\,\bigl(1-(1/s_0+\ii\delta)t\bigr)^{-a}\\
\bigl(s_0+t+\ii\delta\bigr)^{-b}&=&
\bigl(s_0(1+(1/s_0-\ii\delta)t)\bigr)^{-b}=
|s_0|^{-b}\,e^{-\ii\pi b}\,\bigl(1+(1/s_0-\ii\delta)t\bigr)^{-b}\\
(s_0+\ii\delta)^{-2c}&=&|s_0|^{-2c}\,e^{-2\pi\ii c},
\end{eqnarray}
as well as $e^{-2\pi\ii c}/|s_0|^{2c+1}=-(1/s_0-\ii\delta)^{2c+1}$ and
$\Gamma(z)\Gamma(1-z)=\pi/\sin(\pi z)$ we obtain
\begin{equation}
I(\omega,k<0)=\frac{\pi\,\Theta(\omega)\,e^{-\ii\frac{\pi}{2}(2c-1)}}
{\Gamma(a+b-1)}\,
\frac{\omega^{a+b-2}}{v_\mathrm{c}^{a+b-1}}\,
\left(\frac{1}{s_0}-\ii\delta\right)^{2c+1}
\int_0^1 dt\,\frac{t^{2c}\,(1-t)^{a+b-2}}
{\bigl(1-(1/s_0+\ii\delta)t\bigr)^a\,\bigl(1+(1/s_0-\ii\delta)t\bigr)^b}.
\label{eq:res1}
\end{equation}
For $k>0$ we place the cut as shown in Fig.~\ref{fig:contour}.  Performing the
same steps as above we find
\begin{equation}
I(\omega,k>0)=\frac{\pi\,\Theta(\omega)\,e^{-\ii\frac{\pi}{2}(2c-1)}}
{\Gamma(a+b-1)}\,
\frac{\omega^{a+b-2}}{v_\mathrm{c}^{a+b-1}}\,
\left(\frac{1}{s_0}+\ii\delta\right)^{2c+1}
\int_0^1 dt\,\frac{t^{2c}\,(1-t)^{a+b-2}}
{\bigl(1-(1/s_0-\ii\delta)t\bigr)^a\,\bigl(1+(1/s_0+\ii\delta)t\bigr)^b}.
\label{eq:res2}
\end{equation}
We can write \eqref{eq:res1} and \eqref{eq:res2} together as
\begin{equation}
I(\omega,k)=
\frac{\pi\,\Theta(\omega)\,e^{-\ii\frac{\pi}{2}(2c-1)}\,\Gamma(2c+1)}
{\Gamma(a+b+2c)}\frac{\omega^{a+b-2}}{v_\mathrm{c}^{a+b-1}}\,
u^{2c+1}\,F_1\bigl(2c+1,a,b,a+b+2c;u^*,-u\bigr),\quad
u=\frac{2\omega}{v_\mathrm{c}k}+\ii\,\sgn{k}\delta.
\label{eq:resultI}
\end{equation}
Here we have used the integral representation \eqref{eq:integral} of Appell's
hypergeometric function~\cite{ErdelyiHTF1}, which is valid for $1<a+b$.
Analytic continuation in the parameters $a$, $b$, and $c$ then yields
$I(\omega,k)$ for $a+b<2$ and $c>-1/2$. At $K_\mathrm{c}=1$ one finds
$F_1\bigl(2c+1,a,b,a+b+2c;u^*,-u\bigr)=F_1\bigl(1,1/2,0,1/2;u^*,-u\bigr)=
1/(1-u^*)$.

In the same way one can show
\begin{equation}
\begin{split}
&\int_{-\infty}^0 dR\int_{-\infty}^\infty dt\,
\frac{e^{\ii(\omega t-kR)}}{\bigl(v_\mathrm{c}t-2R-\ii\delta\bigr)^c}\,
\frac{(-\ii)^{a+b+2c}}{\bigl(v_\mathrm{c}t+2R-\ii\delta\bigr)^c}\,
\frac{(2R)^{2c}}{(v_\mathrm{c}t-\ii\delta)^{a+b}}\\
&\quad=
\pi\,\Theta(\omega)\,e^{-\ii\frac{\pi}{2}(2c-1)}\,
\frac{\omega^{a+b-2}}{v_\mathrm{c}^{a+b-1}}\,
\frac{\Gamma(a+b+1)}{\Gamma(2a+2b)}\,
u^{2c+1}\,
F_1\bigl(a+b+1,c,c,2a+2b;u^*,-u\bigr),\quad 
u=\frac{2\omega}{v_\mathrm{c}k}+\ii\,\sgn{k}\delta,
\end{split}
\end{equation}
as well as ($A>0$)
\begin{equation}
\begin{split}
&\int_{-\infty}^0 dR\int_{-\infty}^\infty dt\,
\frac{e^{\ii(\omega t-kR)}}{\bigl(v_\mathrm{c}t-2R-\ii\delta\bigr)^a}\,
\frac{(-\ii)^{a+b+2c}\,e^{AR}}{\bigl(v_\mathrm{c}t+2R-\ii\delta\bigr)^b}\,
\left(\frac{2R}{v_\mathrm{c}t-\ii\delta}\right)^{2c}
=\frac{\pi\,\Theta(\omega)\,e^{-\ii\frac{\pi}{2}(2c-1)}}
{\omega^{2-a-b}\,v_\mathrm{c}^{a+b-1}}\,
\frac{\Gamma(2c+1)}{\Gamma(a+b+2c)}\\
&\hspace{30mm}\times
\biggl(\frac{2\omega}{v_\mathrm{c}k}+\ii\,\sgn{k}\delta\biggr)^{2c+1}\,
F_\mathrm{D}^{(3)}\Bigl(2c+1,a,b,2c,a+b+2c;
\tfrac{2\omega}{v_\mathrm{c}k}-\ii\tfrac{A}{k},
-\tfrac{2\omega}{v_\mathrm{c}k}-\ii\tfrac{A}{k},
-\ii\tfrac{A}{k}\Bigr),
\end{split}
\label{eq:FTLauricella}
\end{equation}
where $F_\mathrm{D}^{(3)}$ denotes Lauricella's hypergeometric function of
three variables (see App.~\ref{sec:hf}). For $a=1/2$ and $b=c=0$
\eqref{eq:FTLauricella} simplifies to $2\ii\sqrt{\pi
  v_\mathrm{c}}\Theta(\omega)/\sqrt{\omega} /(v_\mathrm{c}k-2\omega+\ii
v_\mathrm{c}A)$.

\section{Hypergeometric function of several variables}\label{sec:hf}
Hypergeometric series of several variables were first studied by
Lauricella~\cite{Lauricella93}. They are defined by
\begin{equation}
F_\mathrm{D}^{(n)}(\alpha,\beta_1,\ldots,\beta_n,\gamma;z_1,\ldots,z_n)=
\sum_{m_1,\ldots,m_n=0}^\infty
\frac{(\alpha)_{m_1+\ldots+m_n}\,
(\beta_1)_{m_1}\ldots(\beta_n)_{m_n}}
{(\gamma)_{m_1+\ldots+m_n}}\,
\frac{z_1^{m_1}\cdots z_n^{m_n}}{m_1!\cdots m_n!},\quad |z_i|<1.
\label{eq:series}
\end{equation}
The special cases~\cite{ErdelyiHTF1} $n=1$ and $n=2$ are Gauss hypergeometric
function $F_\mathrm{D}^{(1)}=F(\alpha,\beta;\gamma;z)$, and Appell's
hypergeometric function
$F_\mathrm{D}^{(2)}=F_1(\alpha,\beta_1,\beta_2,\gamma;z_1,z_2)$, respectively.
The function $F_\mathrm{D}^{(n)}$ possesses the Euler-type integral
representation~\cite{ErdelyiHTF1,Exton78}
\begin{equation}
\label{eq:integral}
F_\mathrm{D}^{(n)}(\alpha,\beta_1,\ldots,\beta_n,\gamma;z_1,\ldots,z_n)=
\frac{\Gamma(\gamma)}{\Gamma(\alpha)\,\Gamma(\gamma-\alpha)}
\int_0^1dt\,\frac{t^{\alpha-1}\,(1-t)^{\gamma-\alpha-1}}
{(1-z_1t)^{\beta_1}\cdots(1-z_nt)^{\beta_n}},
\quad\mathfrak{Re}\,\alpha>0,\;\mathfrak{Re}\,(\gamma-\alpha)>0.
\end{equation}
Furthermore the following relations hold~\cite{Lauricella93,ErdelyiHTF1}
\begin{eqnarray}
F_\mathrm{D}^{(n)}(\alpha,\beta_1,\ldots,\beta_n,\gamma;z_1,\ldots,z_n)
\!\!&=&\!\!(1-z_1)^{-\beta_1}\cdots(1-z_n)^{-\beta_n}
F_\mathrm{D}^{(n)}\Bigl(\gamma-\alpha,
\beta_1,\ldots,\beta_{n},\gamma;
\frac{z_1}{z_1-1},\ldots,\frac{z_n}{z_n-1}\Bigr),\qquad\label{eq:FDrelation1}\\
F_1(\alpha,\beta_1,\beta_2,\gamma;1,1)\!\!&=&\!\!
\frac{\Gamma(\gamma)\,\Gamma(\gamma-\alpha-\beta_1-\beta_2)}
{\Gamma(\gamma-\alpha)\,\Gamma(\gamma-\beta_1-\beta_2)}\qquad
\text{for}\;\gamma\neq 0,-1,-2,\ldots\;\text{and}\;
\gamma>\alpha+\beta_1+\beta_2.\label{eq:FDrelation2}
\end{eqnarray}

\section{Properties of $\bs{N_\sigma^>(E,2k_\mathrm{F}+q)}$}
\label{sec:singularities}
In order to analyze the dispersing features and singularities of
\eqref{eq:Npos} we first note that $F_1\bigl(2c+1,a,b,a+b+2c;u^*,-u\bigr)$
possesses singularities at $u=\pm 1$.

Let us first study $N_1^>$. The integrand has singularities at 
\begin{equation}
\text{(i)}\quad E-\Delta\cosh\theta=0,\qquad
\text{(ii)}\quad 2(E-\Delta\cosh\theta)=\pm v_\mathrm{c}q.
\end{equation}
Inserting (i) into (ii) immediately yields a feature at $q=0$.  Using
\eqref{eq:FDrelation1} and \eqref{eq:FDrelation2} one can extract the
$q$-dependence $(v_\mathrm{c}q)^{a+b-2c-1}$ to obtain \eqref{eq:CDWpeak}.  On
the other hand, (i) will be stationary at $\theta\approx 0$. Inserting this
into (ii) directly yields the dispersion relation \eqref{eq:cdispersion}. The
suppression of the dispersing peak for $q<0$ follows from the relative
strength of the singularities at $u=\pm 1$.

In the same way the integrand in $N_2^>$ has singularities at 
\begin{equation}
\text{(i)}\quad E-\Delta\cosh\theta=0,\qquad
\text{(iii)}\quad
2v_\mathrm{s}\bigl(E-\Delta\cosh\theta\bigr)=
\pm v_\mathrm{c}\bigl(v_\mathrm{s}q-2\Delta\sinh\theta\bigr).
\end{equation}
Inserting (i) into (iii) directly yields the dispersion relation
\eqref{eq:sdispersion}. Furthermore, we can rewrite (iii) as
\begin{equation}
\text{(iv)}\quad \frac{E}{\Delta}\mp\frac{v_\mathrm{c}q}{2\Delta}=
\cosh\theta\mp\frac{v_\mathrm{c}}{v_\mathrm{s}}\sinh\theta.
\end{equation}
If and only if $v_\mathrm{c}<v_\mathrm{s}$, the right-hand side in (iv)
becomes stationary at $\theta=\tilde{\theta}=\pm\text{arcosh}
\bigl(v_\mathrm{s}/\sqrt{v_\mathrm{s}^2-v_\mathrm{c}^2}\bigr)$. In principle,
this leads to the relation \eqref{eq:scdispersion} for arbitrary $q$. However,
this dispersing feature only exists when
$-\text{arcosh}\left(\tfrac{E}{\Delta}\right)\le\tilde{\theta}\le\text{arcosh}\left(\tfrac{E}{\Delta}\right)$.
Together with \eqref{eq:scdispersion} this yields the condition $q_0\le |q|$.

Finally, in order to prove \eqref{eq:Nbbssingularity} we use that
$F_\mathrm{D}^{(3)}\bigl(2c+1,a,b,2c,a+b+2c;u_3^*,-u_3,-u_3'\bigr)$ is regular
as $E\rightarrow E_\mathrm{bbs}+$, which directly yields
$\alpha=1-a-b-2c=1-1/(2K_\mathrm{c}^2)$.


\begin{thebibliography}{10}
\bibitem{caveat} In Ref.~[\onlinecite{FradkinHirsch83}] only the half-filled
  case was discussed. At half-filling, the CDW is commensurate and, due to the
  existence of an Umklapp operator, the CDW does not slide. The effective
  field theory is a non-chiral SU(2) Gross-Neveu model. In this regime the
  system is effectively a Mott insulator. Away from half-filling the CDW
  slides as the Umklapp operator is absent (irrelevant), and the effective
  field theory is a chiral SU(2) Gross-Neveu model which is equivalent to a
  U(1) Thirring model with two flavors. This model has a continuous chiral
  symmetry, a consequence of the sliding invariance of the CDW.  This is, in
  fact, the generic description of the low-energy physics of a 1D system with
  a CDW ground state, regardless the microscopic origin of this state.
\end{thebibliography}
\end{document}